\documentstyle[12pt,epsf,epsfig,amsmath]{article}
\newfont{\thiplo}{msbm10 scaled\magstep 2}
\newfont{\gothic}{eufb10 scaled\magstep 2}
\newfont{\unc}{eurb10} 
\newskip\humongous \humongous=0pt plus 1000pt minus 1000pt
\def\caja{\mathsurround=0pt}\def\eqalign#1{\,\vcenter{\openup1\jot \caja
        \ialign{\strut \hfil$\displaystyle{##}$&$ 
        \displaystyle{{}##}$\hfil\crcr#1\crcr}}\,}
\newif\ifdtup

\def\eqright #1\cr{\noalign{\hfill$\displaystyle{{}#1}$}}
\def\eqleft #1\cr{\noalign{\noindent$\displaystyle{{}#1}$\hfill}}
\def\oldreffmt#1{\rlap{[#1]} \hbox to 2\parindent{}}

\def\figfmt#1{\rlap{Figure {#1}} \hbox to 1in{}}

\def\VEV#1{\left\langle #1\right\rangle}

\def\sectioneq{\def\theequation{\thesection.\arabic{equation}}{\let
\holdsection=\section\def\section{\setcounter{equation}{0}\holdsection}}}%

\newcounter{holdequation}

\def\auto{\eqno(\refstepcounter{equation}\theequation)}
\def\begineq #1\endeq{$$ \refstepcounter{equation}\eqalign{#1}\eqno
	(\theequation) $$}
\def\contlimit{\,{\hbox{$\longrightarrow$}\kern-1.8em\lower1ex
\hbox{${\scriptstyle (a\rightarrow0)}$}}\,}
\def\centeron#1#2{{\setbox0=\hbox{#1}\setbox1=\hbox{#2}\ifdim
\wd1>\wd0\kern.5\wd1\kern-.5\wd0\fi
\copy0\kern-.5\wd0\kern-.5\wd1\copy1\ifdim\wd0>\wd1
\kern.5\wd0\kern-.5\wd1\fi}}
\def\centerover#1#2{\centeron{#1}{\setbox0=\hbox{#1}\setbox
1=\hbox{#2}\raise\ht0\hbox{\raise\dp1\hbox{\copy1}}}}
\def\centerunder#1#2{\centeron{#1}{\setbox0=\hbox{#1}\setbox
1=\hbox{#2}\lower\dp0\hbox{\lower\ht1\hbox{\copy1}}}}
\def\lsim{\;\centeron{\raise.35ex\hbox{$<$}}{\lower.65ex\hbox
{$\sim$}}\;}
\def\gsim{\;\centeron{\raise.35ex\hbox{$>$}}{\lower.65ex\hbox
{$\sim$}}\;}
\def\st#1{\centeron{$#1$}{$/$}}
\def\super#1{\ifmmode \hbox{\textsuper{#1}}\else\textsuper{#1}\fi}
\def\textsuper#1{\newcount\holdspacefactor\holdspacefactor=\spacefactor
$^{#1}$\spacefactor=\holdspacefactor}
\def\getcite#1,{\advance\citenumber by1
\def\getcitearg{#1}\def\lastarg{@}
\ifnum\citenumber=1
\ref{#1}\let\next=\getcite\else\ifx\getcitearg\lastarg\let\next=\relax
\else ,\ref{#1}\let\next=\getcite\fi\fi\next}
\def\pom{{\rm P\kern -0.53em\llap I\,}}
\def\spom{{\rm P\kern -0.36em\llap \small I\,}}
\def\sspom{{\rm P\kern -0.33em\llap \footnotesize I\,}}

\relax
\def\contlimit{\,{\hbox{$\longrightarrow$}\kern-1.8em\lower1ex
\hbox{${\scriptstyle (a\rightarrow0)}$}}\,}
\def\upon #1/#2 {{\textstyle{#1\over #2}}}
\relax
\renewcommand{\thefootnote}{\fnsymbol{footnote}} 

\def\mainhead#1{\setcounter{equation}{0}\addtocounter{section}{1}
  \vbox{\begin{center}\large\bf #1\end{center}}\nobreak\par}
\sectioneq
\def\subhead#1{\bigskip\vbox{\noindent\bf #1}\nobreak\par}

\def\til#1{\centeron{\hbox{$#1$}}{\lower 2ex\hbox{$\char'176$}}}
\def\tild#1{\centeron{\hbox{$\,#1$}}{\lower 2.5ex\hbox{$\char'176$}}}
\def\sumtil{\centeron{\hbox{$\displaystyle\sum$}}{\lower
-1.5ex\hbox{$\widetilde{\phantom{xx}}$}}}

\def\q{\unc q}

\begin{document} 

\begin{titlepage} 

$~$

\vspace{1in}

\begin{center} 
{\Large \bf Could a Weak Coupling Massless SU(5) Theory Underly the Standard Model S-Matrix?}

\medskip

Alan. R. White\footnote{arw@hep.anl.gov }
 
\vskip 0.6cm

\centerline{Argonne National Laboratory, Il 60439, USA.}
\vspace{0.5cm}

\end{center}

\begin{abstract} 
The unitary Critical Pomeron 
connects to a unique massless left-handed SU(5) theory that, remarkably, might provide an unconventional underlying
unification for the Standard Model. Multi-regge theory suggests the existence of  
a {\it bound-state high-energy S-Matrix} 
that replicates Standard Model states and interactions via massless fermion anomaly dynamics. Configurations of anomalous wee gauge boson reggeons play a vacuum-like role. All particles, including neutrinos, are bound-states with
dynamical masses {\it (there is no Higgs field)} that are formed (in part) by anomaly poles. The contributing zero-momentum chirality transitions break the SU(5) symmetry to vector SU(3)$\otimes$U(1) in the S-Matrix. 
The high-energy interactions are vector reggeon exchanges accompanied by 
wee boson sums {\it (odd-signature for the strong interaction and 
even-signature for the electroweak interaction)} that strongly enhance couplings. The very small SU(5) coupling,
$\alpha_{\scriptscriptstyle QUD} \raisebox{0.5mm}{
\centerunder{${\scriptstyle < }$}{${\scriptstyle \sim } $}}$~$1/120$, should 
be reflected in small (Majorana) neutrino masses.
A color sextet quark sector, still to be discovered, 
produces both Dark Matter and Electroweak Symmetry Breaking.
Anomaly color factors imply this sector
could be produced at the LHC with {\it large cross-sections}, and would be definitively identified in double pomeron processes.
\end{abstract} 

\vspace{0.3in}

\centerline{Contributed  to the Proceedings of the Gribov-80} 
\centerline{Memorial Workshop {\it \{without the Appendix\}}.}

\renewcommand{\thefootnote}{\arabic{footnote}} \end{titlepage} 

\section{\large \bf The Critical Pomeron Leads First to  Electroweak Symmetry Breaking and Dark Matter, Then to Neutrino Masses.}

The Reggeon Field Theory Critical Pomeron
(alone) satisfies all high-energy unitarity constraints\cite{arw1}. Supercritical RFT 
implies\cite{arw1} it occurs via reggeon anomaly interactions in ``~QCD$_S$~'' - QCD with six color triplet quarks plus two color sextet quarks.
SU(3) color produces the right superconducting phase, while 
adding the sextet sector provides the infra-red scaling interactions and 
asymptotic freedom saturation needed for the critical behavior. 

Srikingly, the sextet sector could also solve what are currently regarded 
as the two most fundamental problems of particle physics. Sextet 
``pions'' provide an effective Higgs sector that produces electroweak symmetry breaking, while stable sextet ``neutrons'' provide Dark Matter. Anomaly color factors imply that the sextet sector, including dark matter and multiple electroweak bosons, 
will dominate high-energy cross-sections and so could be responsible for the dominance of Dark Matter production in the early universe. These cross-sections could also be responsible for the Cosmic Ray spectrum knee. If so, they should be seen at the LHC.

Most extraordinarily, though, consistently adding the electroweak interaction to QCD$_S$ requires a unique  massless SU(5) theory that might provide an unexpected and novel origin for the Standard Model. I will suggest that the Standard Model may actually be reproducing an anomaly-driven bound-state S-Matrix that contains both the Critical Pomeron and massive neutrinos and which
sits within the SU(5) theory. Small neutrino masses could be direct evidence for the, {\it perforce very small,} SU(5) coupling.

%\vspace{0.1in}
\section{QUD - a Bound-State S-Matrix Theory?}  

Kyungsik Kang and I discovered\cite{arw1} some years ago that, {\it \large uniquely,} 
\begin{center}{\bf QUD\footnote{
Quantum Uno/Unification/Unique/Unitary/Underlying Dynamics} $\equiv~ $ 
SU(5) gauge theory with left-handed couplings
\newline $~~~~$ to ${\bf 5 \oplus 15 \oplus 40 \oplus 45^*}$ massless fermions }
\end{center}
\noindent contains a potential
electroweak symmetry-breaking color sextet doublet,
is anomaly free, and is asymptotically free.
Under $SU(3)_C\otimes SU(2)_L\otimes U(1)$

\newpage {\small $$
\eqalign{&{\bf 5=[3,1,-\frac{1}{3}]^{3}
+[1,2,\frac{1}{2}]^{ 2}~,~~~~~ 15=[1,3,1]+
[3,2,\frac{1}{6}]^{1}+\{6,1,-\frac{2}{3}\}~,} \cr
&{\bf 40=[1,2,-\frac{3}{2}]^{3}
+[3,2,\frac{1}{6}]^{2}+
[3^*,1,-\frac{2}{3}]~+~[3^*,3,-\frac{2}{3}]+
\{6^*,2,\frac{1}{6}\}+[8,1,1]~,}\cr
&{\bf
 45^*=[1,2,-\frac{1}{2}]^{1}+[3^*,1,\frac{1}{3}]
+[3^*,3,\frac{1}{3}]+[3,1,-\frac{4}{3}]
+[3,2,\frac{7}{6}]^{3}+
\{6,1,\frac{1}{3}\} +[8,2,-\frac{1}{2}]}}
$$}
It was a welcome surprise that, in addition to the sextet quarks \{...\} having the right quantum numbers for sextet ``pions'' to provide the longitudinal components of the electroweak vector bosons, QUD also 
contains QCD$_S$. In fact, both
the triplet quark and lepton sectors, neither of which were asked for, are amazingly close to the Standard Model. There are``almost'' three generations - denoted by superscripts {\bf 1,2,3}.

Very importantly, QUD is real {\it \{vector-like\}} with respect to $SU(3)_C\otimes U(1)_{em}$. Obviously, the $SU(2)_L \otimes U(1)$ quantum numbers are not quite right for the Standard Model, but (also very importantly) the lepton anomaly is correct.
Only after I fully understood the reggeon anomaly dynamics of QCD$_S$, did I realise that QUD could be physically realistic if the same dynamics 
is present. In this case, {\it all elementary leptons and quarks
would be confined and remain massless, with infra-red anomalies dominating the dynamics!!} The Standard Model would have to be an effective theory obtained (in principle) by integrating out the elementary leptons.
 
A priori, the bound-states of a massless field theory are prohibitively
difficult to access. Fortunately,
multi-regge theory provides a key!  As is well-known, infinite momentum wee partons can, in principle, play a vacuum role. 
The multi-regge region involves multiple infinite-momenta that, specifically for QCD$_S$ and QUD, allow 
``universal wee partons''
to play this role in the reggeon diagram construction 
of bound-state amplitudes. Crucially, the infinite-momenta also 
introduce anomaly pole bound-states.

\section{The QUD S-Matrix}

Because of it's uniqueness, QUD is either right or wrong, in it's entirety.
If it fails to reproduce the Standard Model S-Matrix,
it is necessarily wrong. Although much further development is obviously needed and very many details are missing, the arguments outlined in the following imply that
\begin{enumerate}{\it 
\item{{\bf All elementary fermions are confined.} Infinite-momentum bound-states contain anomaly poles involving zero-momentum chirality 
transitions that produce {\bf ~SU(5) $\to$ SU(3)$\otimes$U(1)$_{em}$~}
symmetry breaking.}

\vspace{0.03in}
\item{Infinite-momentum {\bf interactions are vector boson reggeons,} accompanied by sums of anomalous 
wee gauge bosons - odd-signature for the pomeron and 
even-signature for the electroweak interactions.}

\vspace{0.03in}
\item{The S-Matrix is a {\bf massless fermion anomaly phenomenon,} without corresponding off-shell amplitudes. Although QUD lies in the SU(5) ``conformal window'' and the symmetry is unbroken at large $k_{\perp}$

\vspace{0.03in}
\centerline
 {\bf the S-Matrix has only Standard Model interactions}
\centerline{\bf and a spectrum of 
Standard Model form.}}

\vspace{0.03in}
\item{{\bf There is no Higgs~!!} All particles, including (Majorana) neutrinos, are bound-states with dynamical masses}}
\end{enumerate}

Consequently, QUD could, perhaps, provide an amazingly economic 
underlying unification
for the Standard Model, while also producing neutrino 
masses. Beyond the known generations
and the sextet quark sector that, potentially, solves the 
other outstanding mysteries of dark matter and electroweak symmetry breaking, there is only a lepton-like octet quark sector and a pair of exotically charged quarks. My construction implies that the octet quark sector is buried in all states in an 
infinite-momentum (light-cone) subtraction role that produces 
leptons and hadrons in Standard Model generations.
Although the physics is both novel and radical, 
it is consistent with all established Standard Model physics 
and explains many puzzles. Unfortunately, the
multi-regge theory that I use to uncover it is so
erudite that general interest may well require, what would
surely be\cite{arw2,arw3},
\begin{center}
{\it {\bf A MAJOR EXPERIMENTAL DISCOVERY,} i.e. the LHC  
\newline observes BIG x-sections for multiple Z's and W's, $N_6$ and $P_6$
pairs and, 
\newline distinctively, $\gamma~\pom \to Z$
{\it together with} $~\pom ~~ \pom
\to ~ZZ,~ WW$  {\it pairs.}$~~~~~~~~~~$}
\end{center}

An immediate major issue is whether, and how, physical scales can be produced by QUD. Unfortunately, I do not yet have the calculational tools to address this issue directly. An infra-red fixed-point implies a very small coupling, with the second-order $\beta$-function giving $\alpha_{\scriptscriptstyle QUD}  \raisebox{0.5mm}{
\centerunder{${\scriptstyle < }$}{${\scriptstyle \sim } $}}$~$1/120$. While this has the great attraction that it might be an immediate explanation for small neutrino masses, it also implies that Standard Model couplings can not be obtained via 
QUD evolution. It is essential, therefore,
that all particles are anomaly-produced bound-states without the off-shell amplitudes needed to invoke evolution. In this case,
$\alpha_{\scriptscriptstyle QUD}$ has no direct physical meaning,
Moreover, the elementary interaction strengths are enhanced, in the 
high-energy S-Matrix, by infinite sums of wee gauge bosons involving
anomaly color factors. Indeed, the larger color factors
for sextet states imply that, at high-energy, this new sector will be produced with cross-sections that are even larger than normal hadronic cross-sections.

\vspace{-0.2in}
\section{Cosmic Rays and Dark Matter} 

Cosmic rays already suggest that new large x-section physics including 
dark matter could appear at the LHC! As shown in Figure 1, 
\begin{figure}[ht]
\centerline{\epsfxsize=3.4in\epsfbox{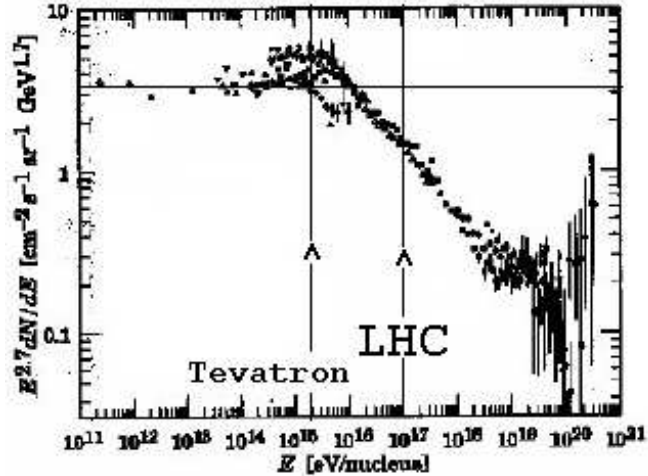}}   
\caption
{The Cosmic Ray Knee}
\end{figure}
the spectrum knee occurs between
Tevatron and LHC energies. It is remarkably 
well-established, yet not understood. Although dark matter
was unknown, a major threshold for
neutral particles, unobserved in 
detectors, was initially suggested ($\sim$ 40 years ago). 
Underestimation of the energy would pile-up events as a ``knee''.
Unbelievably, perhaps, neutral particles must also dominate the x-section far above the knee. But, if the  dark matter x-section is large at the highest LHC energy, a link to the knee is surely inevitable.

{\openup-0.75\jot For the sextet sector, three related effects could produce a knee.
{ \parindent -0.01in 
\begin{enumerate} 
\item{\it Prolific production of electroweak bosons increases $<p_{\perp}>$
dramatically and increases neutrino production
- leading to energy underestimation.}
\item{\it Direct production of sextet neutron dark matter.}
\item{\it Sextet neutrons as incoming cosmic rays (including UHE ?) 
 with a threshold for atmospheric interaction not far below the knee.}
\end{enumerate}
But, can sextet x-sections be large enough 
to dominate the total x-section at the highest energies? 
If so, 
a natural explanation for the existence and dominance of stable dark matter in the universe would be provided.}}

\vspace{-0.2in}
\section{Wee Partons in the Multi-Regge Region}

Multiple regge poles appear in multi-regge limits of multiparticle amplitudes. Most familiar, perhaps, is the triple-regge limit, illustrated in Figure 2(a),
in which $P_1,P_2,P_3 \to \infty$ along distinct light-cones.
In the $ P_3$ 
rest-frame, the regge pole
pions have $\infty$-momentum and
continuation to $Q^2_1=Q_2^2= m_{\pi}^2$
gives the on-shell pion coupling to the pomeron. 
\begin{figure}[ht]
\centerline{\epsfxsize=4in\epsfbox{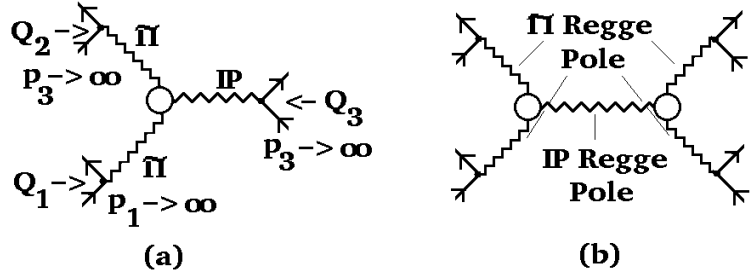}}   
\caption{(a) The Triple-Regge Limit (b) The Di-Triple-Regge Limit}
\end{figure}
More important, for our purposes, is the ``di-triple regge'' 
(DTR) limit in which two triple-regge limits
are separated by a further $\infty$-momentum. Now,
as illustrated in Figure 2(b), 
regge-pole pions can scatter via the pomeron. 
All the pions and
the pomeron  have $\infty$-momentum in some frame, suggesting that
both bound-states ($\pi 's$) and interactions ($~\pom~$) could
appear as reggeon states in which ``universal wee partons'' play a vacuum role. If this can be shown, it amounts to a derivation of (much more than)
the parton model.

In my construction of DTR amplitudes
an initial, cut-off induced, $k_{\perp}$ infra-red divergence produces universal wee gluon reggeons in both QCD$_S$ and QUD, as illustrated in Figure 3. The wee gluon reggeons have 
opposite sign color and space parities and so are ``anomalous''.
\begin{figure}[ht]
\centerline{\epsfxsize=4.9in\epsfbox{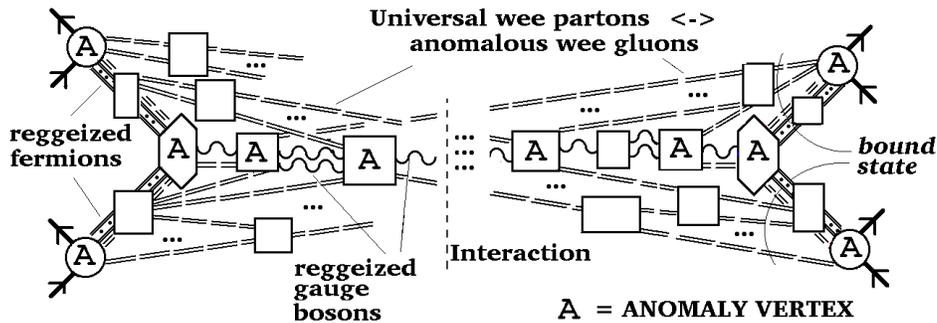}}   
\caption
{A Typical Initial QCD$_S$/QUD DTR Amplitude}
\end{figure}
They couple via anomaly vertices that are chirality violating and contain {\it infra-red chirality transitions}
of massless quarks. Such vertices, {\it necessarily}, involve more than one DTR reggeon channel. Moreover, in combined $\infty$-momentum and
small $k_{\perp}$ limits,
they reduce to {\bf anomaly poles} 
that provide both {\bf bound-state particles} and the {\bf vertex factorization} of the wee gluons.
After outlining the origin of reggeon anomaly vertices, I will discuss QCD$_S$ 
briefly and then discuss QUD 
in more detail.

That Standard Model states and interactions emerge in QUD (in my construction) as the complexity of the wee partons increases, is what has to be demonstrated. The wide range of  scales and interactions in the Standard Model has to be a consequence of the build-up of the (infinite) variety of wee parton anomaly vertices that couple the interactions in the distinct DTR channels.
The arguments that follow will uncover, at best, only the simplest components of what is surely a truly complicated, if beautifully elaborate, phenomenon.

\section{Anomalies and Anomaly Poles}

Reggeon diagrams are generated when
large light-cone momenta are routed through feynman diagrams
so that internal particles are
maximally close to mass-shell, while also having large relative rapidities.
Internal particles with finite relative rapidity generate reggeon interaction
vertices
and, in a gauge theory, fermion loop reggeon vertices include triangle anomalies.
Because a four-dimensional interaction 
is involved, the anomalies occur only in 
special vertices coupling reggeon channels with distinct light-cone momenta
(such as appear in the DTR limit). Included are axial-vector/vector/vector triangle
diagrams ${\scriptstyle T^{\scriptscriptstyle AVV}}$ that, in both QUD
and massless QCD$_S$, must be defined as the zero fermion mass limit of massive
reggeon diagrams. 

At first sight, chirality is conserved in zero mass triangle diagrams, implying 
${\scriptstyle T^{\scriptscriptstyle AAA} ~=~ T^{\scriptscriptstyle AVV} ~= ~T^{\scriptscriptstyle RRR} + T^{\scriptscriptstyle LLL}}$
- producing a conflict between the axial-vector anomaly and  vector
current conservation. 
However, the regularization 
of $\gamma_5$ amplitudes is a major problem. Fortunately, for our purposes, it can be shown that vector current conservation plus the axial anomaly implies unique massless infra-red anomaly pole chiral amplitudes.
\begin{figure}[ht]
\centerline{\epsfxsize=5in\epsfbox{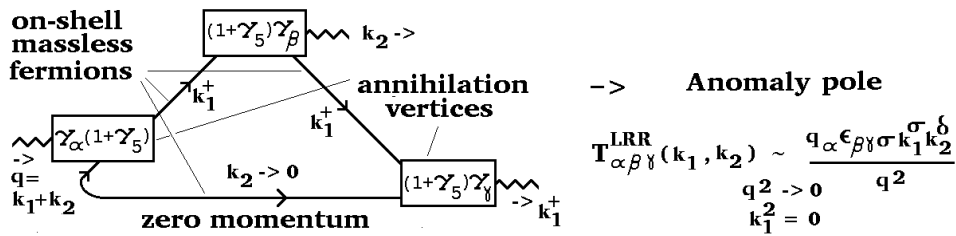}}   
\caption
{Zero Momentum Chirality Violation in $T^{\scriptscriptstyle LRR}$.}  
\end{figure}

In infra-red limits, pseudoscalar anomaly poles appear similiarly in chirality-violating and non chirality-violating amplitudes via the triangle singularity. When there is a chirality transition, as illustrated in Figure 4, the pseudoscalar pole can be a  chiral Goldstone boson.
Infinite-momentum chirality transitions can also occur as part of
a Pauli-Villars subtraction. 

\vspace{-0.2in}
\section{Massless QCD$_S$}

The structure of DTR amplitudes in massless QCD$_S$ is summarized, in a first approximation, in Figure 5. 
\begin{figure}[ht]
\centerline{\epsfxsize=5.5in\epsfbox{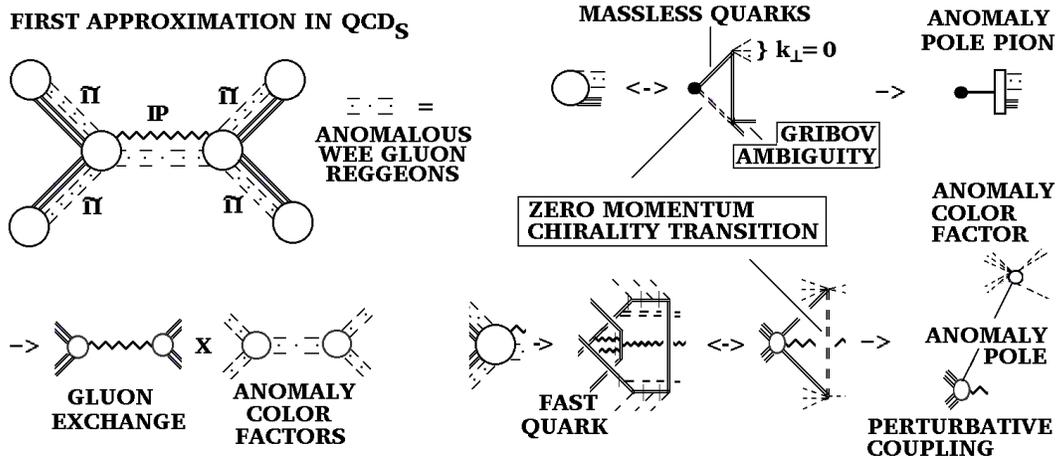}}   
\caption{A DTR Amplitude for Pion Scattering in Massless QCD$_S$}
\end{figure}
Anomalous wee gluon reggeons appear in both the pions and the pomeron via 
anomaly vertices involving zero-momentum quark chirality transitions
(and, for the pion, a longitudinal gluon exchange allowed by the Gribov
ambiguity). As illustrated, the full pion coupling to the pomeron also contains a perturbative
coupling of the dynamical quark and gluon reggeons that are involved.  
The wee gluons lie in an SU(2) color subgroup, but 
their combination with the dynamical reggeons  
produces a color zero projection in each channel. This results in SU(3)
color zero states when the SU(2) subgroup 
is randomized (averaged over) within SU(3), as we briefly discuss later, via the Critical Pomeron.
 
\parindent 0.05in {\openup-0.6\jot
\begin{itemize}{\it
\item{The bound-states are triplet and sextet (pseudoscalar) mesons and 
\newline {\it (with an extra quark reggeon)} triplet and sextet baryons. There are}
\item{{\bf NO} pseudoscalar anomaly poles producing hybrid sextet/triplet states,} 
\item{{\bf NO} glueballs, {\bf NO} BFKL pomeron, and {\bf NO} odderon.}
\item{The Critical Pomeron occurs as a factorized regge pole, plus 
triple pomeron interactions - consistent with the parton model.}}
\end{itemize}}

If sextet pions become the longitudinal components of massive electroweak bosons, sextet nucleons 
are the only new states. (The $\eta_6$ aquires an electroweak scale mass by mixing with the pomeron.)
That sextet quarks have zero current mass implies the $N_6$ neutron is stable. (Electric charge makes the  
$P_6$ proton heavier - in contrast to the triplet sector.) The very strong, very short range, QCD self-interaction implies the $N_6$'s could form DARK MATTER ``clumps''. They will only interact with normal matter at
ultra high energies.  $N_6$ production will dominate the high-energy cross-sections responsible for early universe stable matter formation, and will also explain the Cosmic Ray knee!

These results are at variance with conventional expections for high-energy QCD.
There are fewer states (than requiring just confinement and chiral symmetry breaking) and the interaction is simpler. Both features are 
strongly suggested by experiment! 
Although the anomaly dynamics appears to require the quarks to be massless, effective quark masses that do not disturb the dynamics are produced by 
embedding QCD$_S$ in QUD. 

\vspace{-0.15in}
\section{QUD Reggeon Diagrams - the Massless Limit}

The chirality transitions in QCD$_S$ do not conflict with the vector gauge symmetry. In QUD, more fundamentally, they produce a dynamical breaking of the non-vector part of the gauge symmetry.
In the following construction they are a direct consequence of the zero fermion mass limit and so they retain the initial mass symmetry breaking. In the physical S-Matrix, they should be 
dynamical and randomized via the Critical Pomeron.

I start with masses for all reggeons 
and a $k_{\perp}$ cut-off $\lambda_{\perp}$. How the masses and cut-off are removed is crucial in resolving the (light-cone) Gribov ambiguity and, in effect, produces the wee partons of the massless theory.
A combination of {\bf 24}
and {\bf 5$\oplus$5$^*$} scalar VeV's is needed to give masses to all the fermions.
This also identifies particle/antiparticle pairs and so determines possible chirality transitions. For the gauge bosons, using
only {\bf 5$\oplus$5$^*$} 
VeVs ensures a smooth massless limit (via complementarity).
\parbox{1.2in}{
%$~$ \newline 
\epsfxsize=1.1in \epsffile{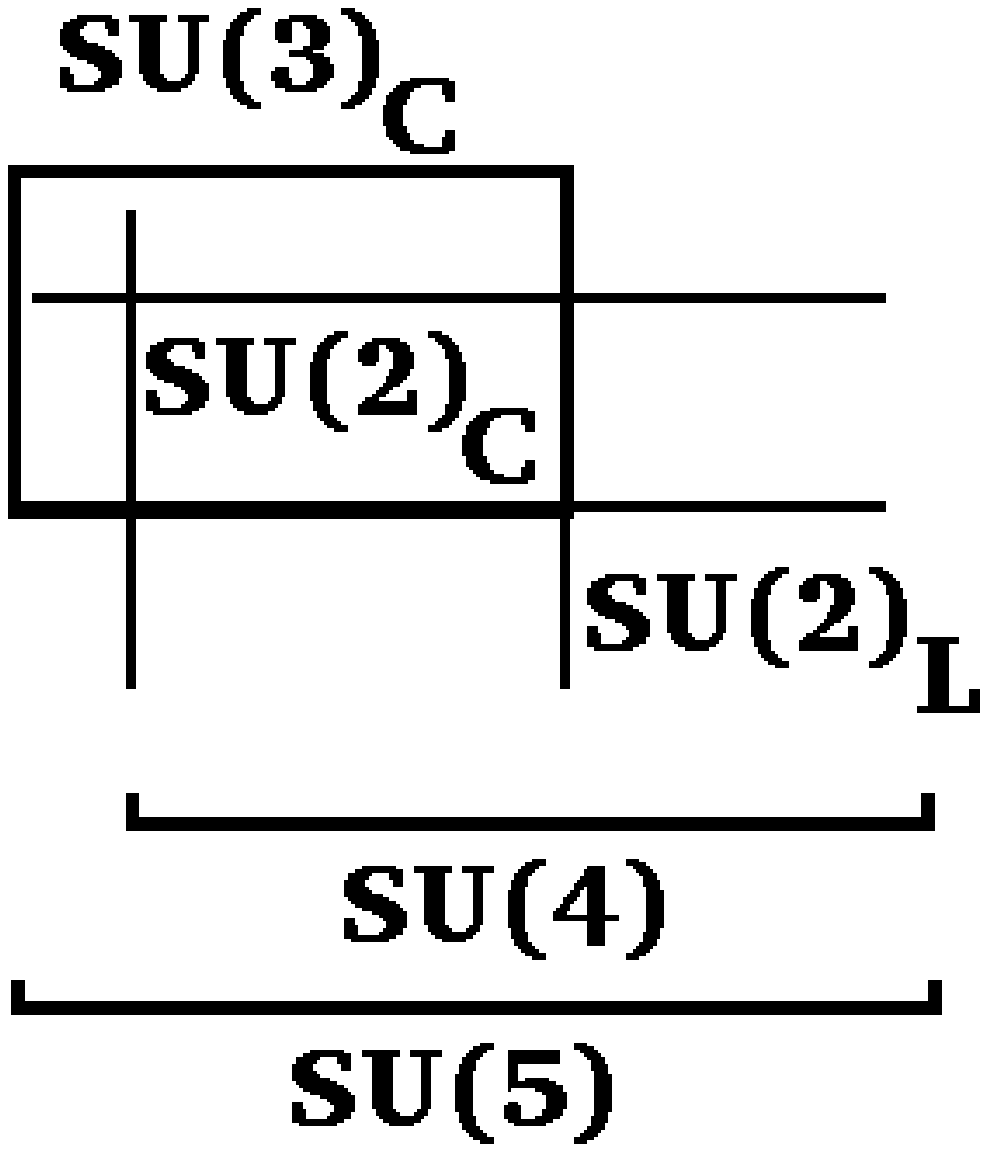}}\parbox{4.5in}{\begin{enumerate} 
\item{{\it I decouple fermion mass scalars first, leaving 
chirality 
transitions that break SU(5) to}
{\bf ~SU(3)$_C\otimes$U(1)$_{em}~$} {\bf in anomaly vertices only.}}
\item{\it I decouple gauge boson scalars successively, giving
global symmetries}
\centerline{${\bf 
\rightarrow ~SU(2)_C~, ~~\rightarrow~  SU(4)~,} ~~~
 \lambda_{\perp} \to \infty~,~~ {\bf \rightarrow }~~{\bf SU(5)}
$}\end{enumerate}}
\newline Closely related to the weak coupling of QUD, the last scalar to be removed is asymptotically free, allowing $\lambda_{\perp} \to \infty$ before the SU(5) limit.
This is essential for obtaining the Critical Pomeron and for the emergence of Standard Model generations via infinite-momentum color octet contributions.

In non-anomaly reggeon diagrams, the exponentiation of reggeization divergences
leaves only infra-red finite, global color zero, interaction kernels. 
{\bf Crucially}, {\it $\lambda_{\perp}$ implies fermion loop vertices,
{\bf including anomaly vertices,} do not satisfy Ward identities producing $k_{\perp} = 0$ zeroes}. Consequently, 
the  exponentiation of divergences goes well beyond reggeization, {\it particularly for
left-handed bosons.}

\vspace{0.03in}
\noindent {\bf Parity non-conservation} allows fermion loop vertices that exponentiate all left-handed bosons contributions that would be part of, or are accompanied by, the anomalous wee boson divergences that we discuss
next.

\vspace{-0.2in}
\section{SU(2) Color Restoration and Anomalous Wee Gluons}

With a $k_{\perp}$ cut-off, {\bf vector} SU(2)$_C$  produces a {\it non-exponentiating} divergence involving
the {\bf anomalous wee gluons} of Figure 3, which are

\vspace{0.05in} 
\centerline{\it  $I=0$ sets of massless gluon
reggeons, with $k_{\perp}$'s}

\centerline{\it scaled to zero and color parity 
$C\neq$ {\Large $\tau$}= signature.}
\noindent 
The anomalous color parity implies that the {\it vector} wee gluons  
couple only via anomaly vertices with chirality transitions. For SU(2), only {\Large $\tau$} $= -C =-1$ is possible  
($\leftrightarrow 3, 5, ... \infty$ reggeons). Infra-red
fixed-point scaling implies 
that the iteration of $I=0$ kernels reproduces the basic divergence,
with a factorized residue. Also, as for QCD$_S$, anomaly poles in vertices 
connect the divergence in different channels. 
Factoring off the overall divergence
leaves a universal wee gluon component in all reggeon states.

As for the pion anomaly pole in Figure 5, bound-state anomaly poles are also  
present in external vertices. To extract residues, it is necessary to go to an infinite-momentum frame in which the wee gluons carry vanishingly small light-cone momenta orthogonal to the infinite-momentum of the fermions and for the polarizations of the fermions and the wee gluons to be (additionally) orthogonal.  In the process generating the 
pole a reggeon state containing a same chirality physical fermion pair and an anomalous wee gluon component, is coupled to a state containing
only two opposite chirality fermions, one of which is 
unphysical and has zero momentum. 
In effect, there is a zero-momentum shift of the Dirac sea.
By absorbing anomalous wee gluons, a physical fermion makes a symmetry-breaking chirality transition to an unphysical ``hole state'' and so produces
a pseudoscalar Goldstone boson\footnote{At infinite-momentum, an anomaly pole
has physical Goldstone boson couplings.}. The chiral symmetries we discuss next 
do not conflict with the SU(2)$_C$ gauge symmetry. Later, when the gauge symmetry
is broken, it will be important that the reggeon state involved has a projection on a (color zero) unbroken symmetry state. 

Via $5\oplus5^*$ chirality transitions, 
reggeon states containing SU(2)$_C$ anomalous wee gluons produce chiral Goldstones ($\pi_C$'s),that are $q\bar{q}~$ ``mesons" or $qq$ and $\bar{q}\bar{q}~$ ``nucleons". The $q$'s are {\bf 3's, 6's,} and {\bf 8's} under SU(3)$_C$. Because the {\bf 8's} are real representations, 
they can not give an SU(3)$_C$ anomaly. However, they contain complex SU(2)$_C$ chiral doublets that produce anomaly poles when only SU(2)$_C$ is restored.
Other reggeon states containing a $\pi_C$ are also 
selected by the wee gluon divergence and will, ultimately, give leptons and SU(5) symmetric reggeon states. To avoid 
fermion loop exponentiation of the anomaly divergence,
the massive gauge boson reggeons in such states must all be vectors 
(i.e gluons or photons).

The leading interaction exchanges are even signature and contain an SU(2)$_C$ singlet massive vector boson accompanied by anomalous wee gluons. As for the pomeron/pion coupling in Figure 5,
the coupling to bound-states contains both an anomaly vertex involving wee gluons 
and a perturbative coupling of dynamical
fermions to the exchanged boson. SU(3)$_C$ massive gluon exchange straightforwardly gives a supercritical pomeron. The gluon 
can also be replaced by a massive $\gamma$ or, after inclusion of the vertices involving bound-state wee gluons that we discuss next, a $~W^{\pm}$ or $~Z^0 $ (using SU(3)$_C \otimes$SU(2)$_L\otimes$U(1) quantum numbers). There are also non-leading odd-signature interactions $\gamma~\pom~$, $W^{\pm}~ 
\pom~$ and $Z^0~\pom~$, that will actually provide the physical $\gamma$, $W^{\pm}$, and $Z^0$, after SU(3)$_C$ restoration.

\section{Bound-State Anomalous Wee Gluon Vertices}

Elementary left-handed $W^{\pm}$ and $Z^0$ exchanges, accompanied by wee gluons, are exponentiated to zero via fermion
loop interactions, but 
5$\oplus$5$^*$ chirality transitions provide crucial 
couplings to the ${\pi_C}'s$ ($\sim {\pi_{6}}'s$) involving wee gluons originating from the scattering bound-states.
In an appropriate $\infty$-momentum frame, the vertices can be evaluated 
via anomaly pole contributions, as illustrated in Figure 6.
\begin{figure}[ht]
\centerline{\epsfxsize=5in\epsfbox{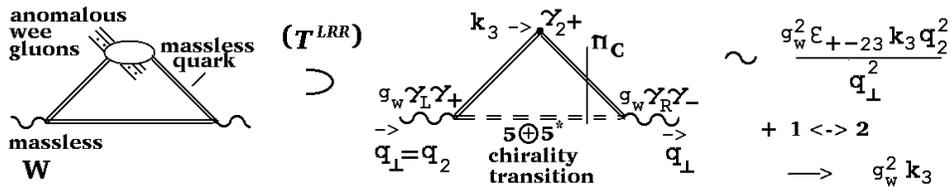}}   
\caption{Vector Boson Mass Generation}
\end{figure}
They provide a mass 
$M_W^2$ ($\sim g_W^2 \int dk k$ - a wee gluon integral multiplied by an electroweak scale determining sextet anomaly factor) that survives the SU(5)
symmetry restoration, while also providing a chirality transition and
sextet flavor quantum number that prevents the exponentiation. The perturbative coupling of the $W^{\pm}$ and $Z^0$ is retained, however.

The {\bf 24} chirality
transitions also provide very important wee gluon vertices, as illustrated in Figure 7. 
\begin{figure}[ht]
\centerline{\epsfxsize=5in
\epsfbox{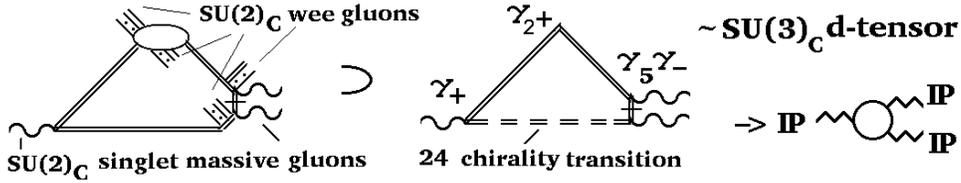}}   
\caption{A Triple Pomeron Vertex}
\end{figure}
The SU(3) d-tensor shown gives a triple pomeron vertex 
with the symmetry properties needed to produce the Critical Pomeron. The wee gluons in the scattering states provide the orthogonal $\gamma$-matrices needed to generate a $\gamma_5$.
%In particular,a reggeon state with a color zero projection and finite %perturbative interactions (when the symmetry breaking is averaged over) connects to 

\vspace{-0.2in}
\section{SU(4) Color Restoration} 

As SU(4) color is restored, a vector-coupling boson $\gamma_4$ that is a singlet under both SU(2)$_C$ and SU(2)$_L$(a linear combination of the $\gamma$ and  
a massive gluon), becomes massless. The $W^{\pm}$ and $Z^0$ survive and retain the mass discussed above, while 
other gauge bosons that become massless are left-handed and their reggeon 
amplitudes are amongst those exponentiated to zero by fermion loop interactions. (They survive, however, in interaction kernels.)
{\large $\gamma_4$} pairs are even signature and have a scaling interaction 
that also exponentiates amplitudes to zero via fermion loop interactions, {\it except} when  
the pairs couple via anomaly vertices ($e^{-\infty} \rightarrow 1 - e^{-\infty}$)
and combine with the anomalous wee gluons. Consequently, 
as reggeon states, bound-states now contain two fermion pairs produced by anomaly poles and accompanied by  odd-signature {\bf anomalous wee bosons}, which are 
\newline $\raisebox{-2mm}{$\hbox{\it ~~~~~~~SU(2)$_C$ anomalous wee gluons plus $\gamma_4$ pairs
(each with $k_{\perp} =0 $).}$}$

\vspace{0.05in}
The anomaly vertices coupling to $\gamma_4$ pairs are produced by
\begin{enumerate}
\item{ {\bf lepton pairs - } 
{\it (1,~2,~$\frac{1}{2}$)/(1,~2,~-$\frac{1}{2}$) chiral symmetry plus} {\bf 24} {\it chirality transitions
gives anomaly pole}  {\bf pseudoscalars} $~\pi_L^{\pm,0}$}

\vspace{0.05in}
\item{{\bf SU(2)$_C$ singlet octets - } {\it (8,~1,~1)/(8,~2,~-$\frac{1}{2}$) chiral symmetry plus {\bf 5$\oplus$5$^*$} chirality transitions gives
anomaly pole} {\bf pseudoscalars} $~\eta_8^{\pm,0}$}
\end{enumerate}

The new Goldstone boson states explicitly break the SU(4) gauge symmetry; but this is already broken by the fermion loop removal of left-handed reggeons. In place of the full symmetry, reggeon states must have SU(4) singlet projections that, after the randomizing of the symmetry breaking within SU(5) (via the Critical Pomeron), will give finite amplitudes.
The fermion states corresponding to SU(4) singlet
reggeon states are then

\vspace{0.1in}  \centerline
{{\bf LEPTONS ~-~} $\pi_L~ + ~\pi_8$
~+~ elementary lepton $\to$ 3 generations.}

\vspace{0.05in}
\centerline{{\bf ~MESONS ~-~} $\pi_{3,6}~+~ \eta_8~$, ~~ {\bf BARYONS} $~\leftrightarrow~$ additional quark.}
\noindent Note that both leptons and hadrons have {\it octet quark} components.

The leading interaction is the even signature supercritical pomeron - the remaining massive SU(4) singlet vector reggeon plus {\it odd-signature
wee bosons.} Interactions involving electroweak 
exchanges are present, but do not have the SU(3)$_C$ singlet projection needed to  survive the SU(5) restoration. The odd-signature exchanges $\gamma~\pom~$, $~W^{\pm}~ 
\pom~$, and $Z^0~\pom~$, accompanied by {\it even-signature
wee bosons,} will
provide the physical interactions.

\section{SU(5) Color Restoration} 

As the remaining SU(4) singlet vector becomes massless, the supercritical
pomeron becomes Critical and, simultaneously, the photon becomes massless. 
The massless photon is, therefore, an odd-signature partner of the even signature Critical Pomeron. There is no ``triple-photon'' vertex and the photon does not have the anomaly couplings to hadrons that make the pomeron interaction so much stronger. We assume that the Critical Pomeron is
analagous to a second-order phase transition phenomenon in that the 
transition randomizes the symmetry breaking. Critical amplitudes can, nevertheless, be obtained from supercritical amplitudes in which order parameters explicitly breaking the symmetry are introduced and then removed. Most likely, the symmetry breaking is randomized in all sets of reggeon diagrams connecting anomaly vertices. As discussed elsewhere\cite{arw1},
the $k_{\perp}$ cut-off is an order parameter that must be removed first, if the restoration of SU(3)$_C$ is to be sufficient to give the critical behavior.

The new massless vector (carrying zero $k_{\perp}$ via a divergence) also joins odd-signature wee bosons  to give {\bf even-signature anomalous wee bosons} that have SU(3)$_C$ color zero and combine with
$\gamma~$, $W^{\pm}$, and $Z^0$ exchanges to give the SU(5) singlet projections
that are randomized to provide the physical electroweak interaction. 

Because the $\pi_8$ and $\eta_8$ combine to form a real octet SU(3)$_C$
representation {\bf ($\Pi_8$)}, the
{\bf 
infra-red octet anomaly poles cancel in all amplitudes.}
However, the $\lambda_{\perp} \to \infty$ limit, taken 
before the 
SU(5) limit, introduces an $\infty$-momentum octet anomaly contribution (as a companion to the canceling infra-red anomaly pole) in 
vertices coupling the SU(3)$_C$ component
of the {\it even-signature wee bosons}. 
Via a Ward identity, the anomaly couples like an $\infty$-momentum $\Pi_8$ anomaly pole so that, in effect, octet quark anomaly poles coupled to SU(3)$_C$ wee gluons provide an $\infty$-momentum
contribution in all bound-states and interaction vertices.

In reggeon bound-states the infra-red dynamical fermion reggeons, that couple perturbatively to the dynamical exchanged vector bosons, must combine with an 
$\infty$-momentum ${\Pi_8}$ pair and adjoint representation anomalous wee bosons to give an SU(5) singlet projection. In the corresponding fermion states, this requires that  
three elementary fermion reggeons, two of which are produced by an anomaly pole, 
combine to provide the $SU(2)\otimes U(1)$ representations
\newline$~$
\centerline{${\bf (2,-\frac{1}{2})_L ~},~~ or ~{\bf (2,\frac{1}{2})_R ~}, ~~ or
 ~{\bf ~~(1,1)_L ~}, ~~or ~{\bf (1,-1)_R} $}

\noindent Consequently, leptons and hadrons form Standard Model generations.

\section{Physical Leptons}

The SU(3)$\times$SU(2)$_L \times$U(1) fermion 
reggeons that combine with SU(5) adjoint wee bosons to form physical leptons are
\begin{itemize}
\item{\it $(e^-,\nu) ~\leftrightarrow
~(1,2,-\frac{1}{2}) \times \pi_L^0 \times [\Pi_8]_{\infty}$ 
\newline {  $~~~~~~~~~ ~\leftrightarrow ~~
(1,2,-\frac{1}{2}) \times (1,2,-\frac{1}{2}) (1,2,\frac{1}{2}) \times [(8,1,1)(8,2,-\frac{1}{2})]_{\infty}$
\newline ${\bf ~~~~~~~~~
\leftrightarrow~~ SU(5)~ adjoint~-~ 45^*\times45^*\times 5\times[40\times45^*]}$}
}
\item{\it $(\mu^-,\nu) ~\leftrightarrow~
(1,2,\frac{1}{2}) \times \pi_L^- \times [\Pi_8]_{\infty}$ 
\newline {$ ~~~~~~~~~ ~\leftrightarrow ~~
(1,2,\frac{1}{2})\times  (1,2,-\frac{1}{2})  (1,2,-\frac{1}{2}) 
\times [(8,1,1)(8,2,-\frac{1}{2})]_{\infty}$}
\newline {${\bf ~~~~~~~~~
\leftrightarrow~~ SU(5)~ adjoint ~-~
5\times45^*\times45^*\times[40\times45^*]}$}}
\item{\it $(\tau^-,\nu) ~\leftrightarrow~ 
(1,2,-\frac{3}{2}) \times \pi_L^+ \times [\Pi_8]_{\infty}$  
\newline {$ ~~~~~~~~~ ~\leftrightarrow~~
(1,2,-\frac{3}{2}) \times (1,2,\frac{1}{2})  (1,2,\frac{1}{2}) 
\times [(8,1,1)(8,2,-\frac{1}{2})]_{\infty}$}
\newline {
$ {\bf ~~~~~~~~~\leftrightarrow~~ SU(5)~ adjoint~ -~ 
40\times5\times5\times[40\times45^*]}$}}
\end{itemize}
In principle, $e^+,~\mu^+$ and $\tau^+$ can be obtained via charge conjugation,
once elementary antiparticles are explicitly identified. Neutrinos will necessarily be Majorana fermions. Chirality transitions connect
the constituents of neutrino and anti-neutrino candidates and there is no quantum number that would prevent anomalous wee boson vertices 
from generating mass terms involving left-handed neutrinos and right-handed antineutrinos.

\vspace{-0.1in}
\section{The Bound-State Mass Spectrum}

I have outlined my understanding of only a small part of the structure of
DTR amplitudes in QUD. It is obviously a major challenge to develop
my outline into a calculable reggeon diagram formalism that would explicitly
provide scales and masses.
Wee gluon anomaly vertices need to be catalogued, as a starting point. They will mix reggeon states and introduce color factors, with 
related wee gluon distributions needed to determine how many, if any, 
parameters are involved. At this stage, we can only say 
\begin{enumerate}
\item{\it Since $\alpha_{\scriptscriptstyle QUD}$ is so small, perturbative reggeization is
a small effect, reflected only in small masses for the, zero color and charge, neutrinos.}
\item{\it SU(3)$_C$ interactions and masses 
will be enhanced: by anomaly color factors, by the triple pomeron interaction,
and by the  high mass sector.}
\item{{\it Assuming
bound-state fermions have constituent masses, connecting the $\eta_6$ to top production suggests\cite{arw2}}

\vspace{0.05in}
\centerline 
{$~~~~~~~~~~~~~m_{q_6} \sim m_{top} \implies m_{N_6} \sim
500 ~GeV$}}

\vspace{0.05in}
\item{\it Electromagnetic anomaly factors will enhance charged particle masses, but less strongly. There is no triple photon interaction.}
\item{\it There is no symmetry conflicting 
with the Standard Model spectrum.}
\item{\it CP violation can be
introduced via the anomalies, but is it essential?} 
\end{enumerate}

\section{Potential QUD Virtues - beyond QCD}

Of course, the scientific and aesthetic importance of an
underlying massless field theory for the Standard Model can not be exagerated. In addition,
\begin{itemize}
\item{\it QUD is self-contained, with  only Standard Model Interactions.
It has to be completely right - or else it is completely wrong!}
\item{\it The massless photon partners the ``massless'' Critical Pomeron.}
\item{\it The only new physics still to be discovered
is a high mass sector of the strong interaction: giving {\bf electroweak symmetry breaking, dark matter,
and unification,} without supersymmetry!}
\item{\it {\bf Parity properties} of the strong, electromagnetic, and weak interactions are naturally explained.}
\item{\it Anomaly vertices mix the reggeon states.  
Color factors could produce the wide range of Standard Model
scales and masses,
with {\bf small Majorana neutrino masses} due to the very  
small QUD coupling.}
\item{\it Despite the underlying SU(5) symmetry, there is {\bf no proton decay.}}
\item{\it Particles and fields are truly distinct, with physical hadrons and leptons having equal status. Symmetries and masses are S-Matrix properties. There are no off-shell amplitudes and there is {\bf no Higgs field.} }
\item{\it The QUD S-Matrix is the only ``non-perturbative'' part of 
quantum field theory needed - with {\bf infinite momentum} physics retaining a {\bf diagrammatic ``parton model'' description.}}
\item{\it Perturbatively, QUD is a massless, asymptotically free,  
fixed-point theory that has no renormalons, and so,
no vacuum energy. Therefore\cite{bh}, it would induce
Einstein gravity with {\bf zero cosmological constant.}}
\end{itemize}

The following Appendix is not included in the Gribov-80 Proceedings contribution.
It is part of an expanded version 
of the foregoing paper that is in preparation.
 
%\newpage

\mainhead{Appendix - The Triangle Anomaly and Anomaly Poles}

Central to the arguments of this paper are reggeon diagram effective vertices 
that contain the fermion triangle diagram anomaly. The triangle diagrams involved originate from much larger loop feynman diagrams in which many internal lines are placed on-shell by the multi-regge limit in which the vertices appear. Because of their special spin and momentum structure, such vertices are an essential component of our construction of bound-state amplitudes in both QCD$_S$ and QUD.
In this Appendix we elaborate the intricate interplay of ultra-violet and infra-red phenomena that we anticipate by describing a variety of properties of the elementary triangle diagram involving axial-vector, vector, and chiral currents. In particular, we will focus on the origin and significance of anomaly poles, as both an infra-red and an ultra-violet phenomenon, including an elaboration of the chirality transitions involved.

Reggeon diagram effective vertices actually contain couplings with much more structure than the local couplings of the diagrams we study here. Nevertheless, we expect the infra-red structure involving on-shell particles to be the same, even though the large momentum region will be more complicated. Therefore, in the paper we assume that infra-red anomaly poles
appear in the same way as in the elementary amplitudes and that the interplay 
with ultra-violet phenomena is essentially the same.

\subhead{A.1 The Definition of $T^{\scriptscriptstyle AVV}$ and $T^{\scriptscriptstyle AAA}$.}

We begin our discussion by considering the amplitude 
$T^{\scriptscriptstyle AVV},$ involving one axial-vector and two vector currents, defined as
$$
T^{\scriptscriptstyle AVV}_{\mu \alpha \beta}(k_1,k_2)~=~
\Gamma_{\mu \alpha \beta}(k_1,k_2)~+~ \Gamma_{\mu \beta\alpha}(k_2,k_1)
\auto\label{amp0}
$$
where
$$
\eqalign{&\Gamma_{\mu \alpha \beta}(k_1,k_2,m)\cr
&~~~~~~~~= {1 \over (2 \pi)^4} \int {  d^4 p~ Tr \{ \gamma_5
\gamma_{\mu} (\st{p} + \st{k}_2 +m )  \gamma_{\alpha}(\st{p} - \st{k}_1 + \st{k}_2 +m ) 
\gamma_{\beta} (\st{p} - \st{k}_1 + m ) \} 
\over  [(p+ k_2)^2 - m^2] [(p - k_1 + k_2)^2 -m^2 ] 
[ (p - k_1)^2 -m^2] }}
\auto\label{tamp}
$$
and the second term results from reversing the direction of the fermion line - thus producing Bose symmetry for the two vector currents. We will also discuss the amplitude $T^{\scriptscriptstyle AAA}$ involving three (identical) axial-vector currents that
would be formally defined by the same integral as $T^{\scriptscriptstyle AVV}$ if (as is not the case) $\gamma_5$'s could be straightforwardly
anti-commuted through the trace numerator. We will discuss this 
issue at some length later, it is very important for our general purpose.

It is very well known that (\ref{tamp}) is
linearly divergent and that a subtraction procedure is necessary to properly define both the $T^{\scriptscriptstyle AVV}$ and $T^{\scriptscriptstyle AAA}$ amplitudes. The subtraction can be carried out by a Pauli-Villars procedure utilising an unphysical fermion whose mass is taken to infinity at the end of calculations. This procedure
directly preserves conservation of the vector currents and, in addition, has other important consequences. As we will illustrate, the subtraction  process
can also be viewed as involving surface terms that generate both the anomaly and
vector current conservation. In this case, the routing of external momenta through the diagram becomes a non-trivial issue and, as is also well-known, it 
is the momentum routing illustrated in Fig.~A.1(a) that
gives vector current conservation. The alternative 
momentum routing illustrated in Fig.~A.1(b) will also play a role.
In our discussion of $T^{\scriptscriptstyle AAA}$  we will use the symmetric
momentum notation illustrated in Fig.~A.1(c), although the definition we use will effectively be a symmetrized sum of amplitudes defined via the momentum routing of Fig.~A.1(a).
\begin{center}
\leavevmode
\epsfxsize=1.9in
\epsffile{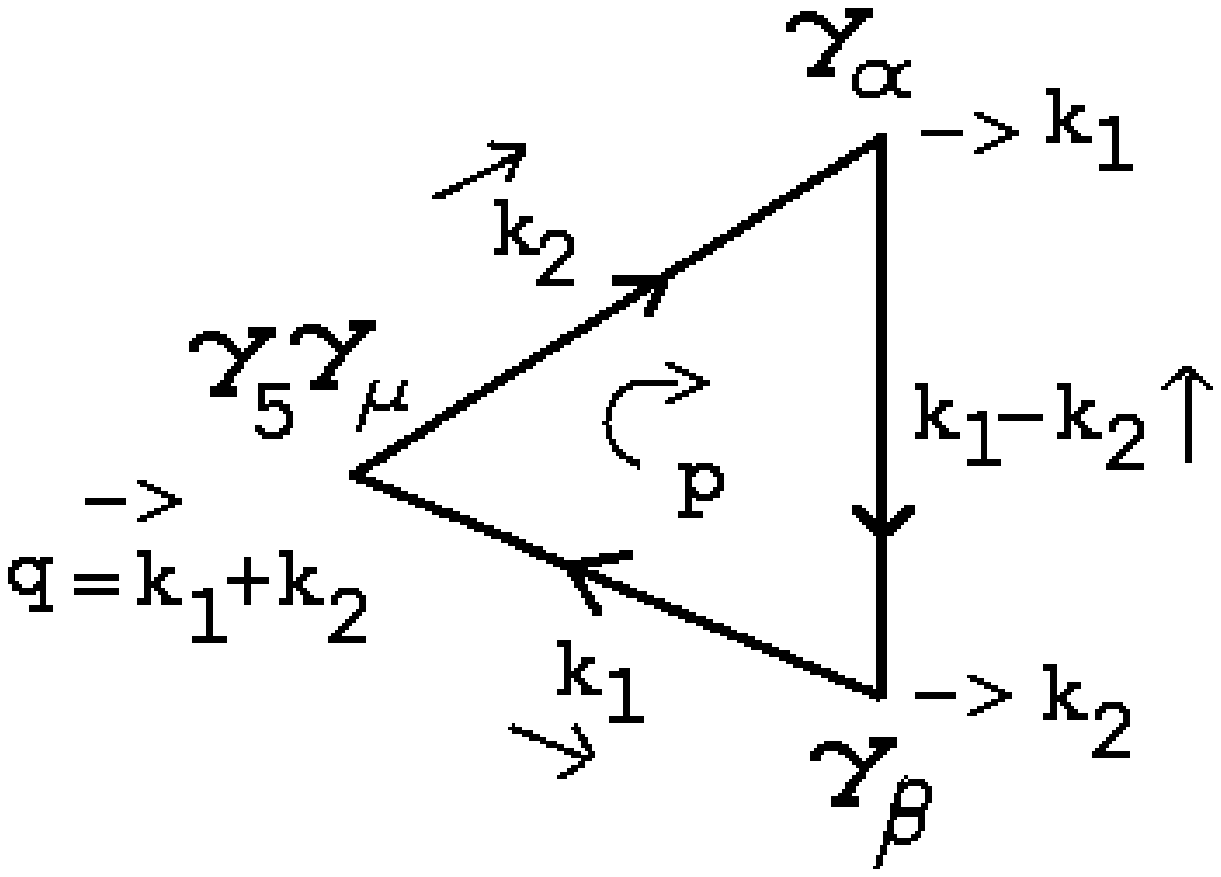}
\hspace{0.2in}
\epsfxsize=1.4in
\epsffile{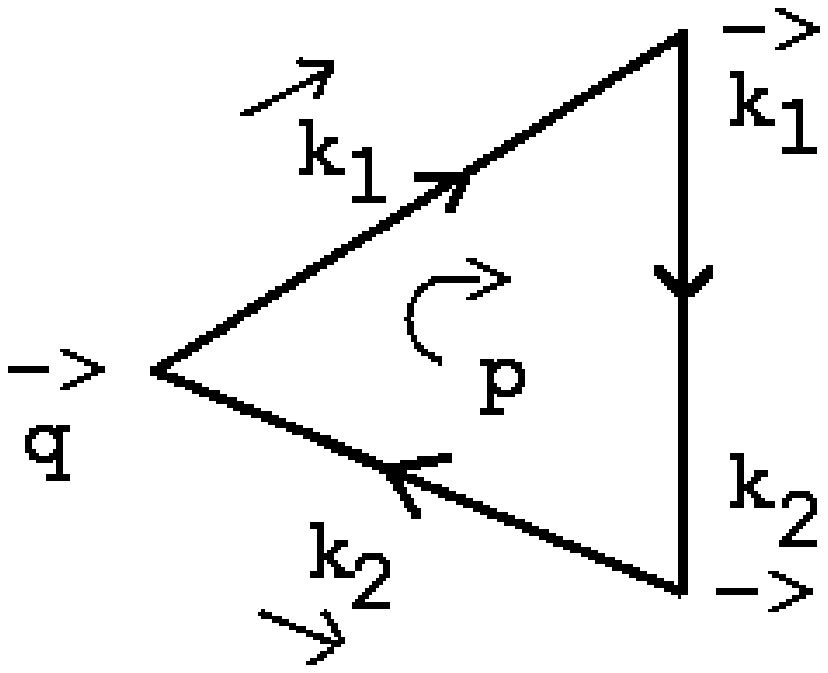}
\hspace{0.35in}
\epsfxsize=1.9in
\epsffile{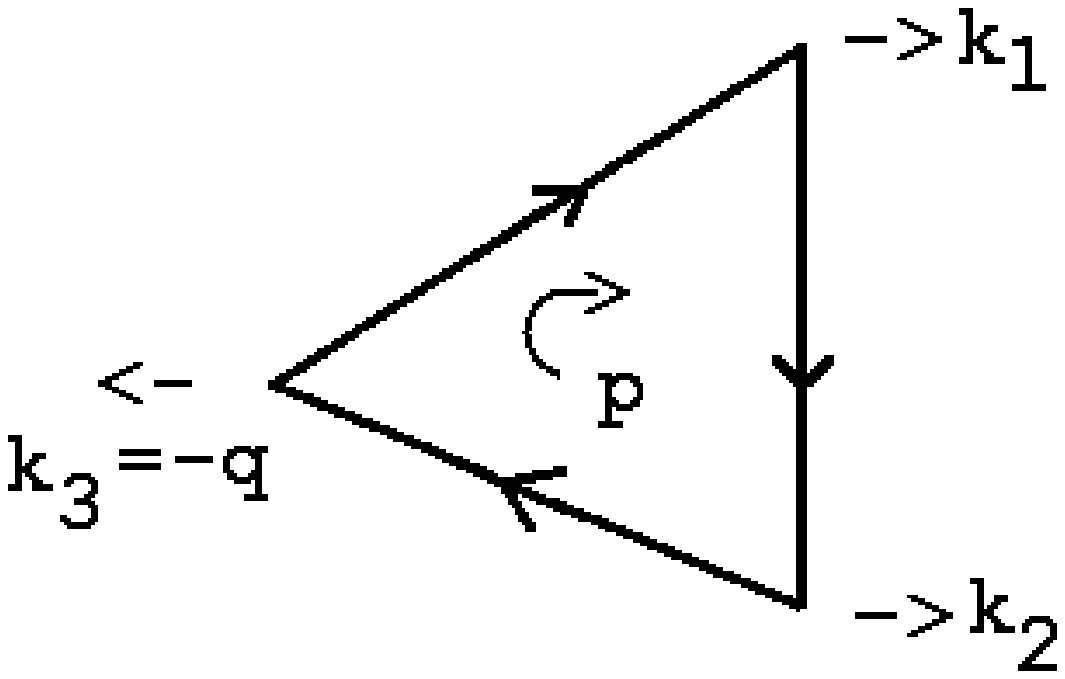}

(a) \hspace{1.6in} (b) \hspace{1.4in} (c)

Fig.~A.1 (a) Notation for $T^{\scriptscriptstyle AVV}$ - the line arrows indicate the fermion direction 
\newline (b) Alternate Momentum Routing for $T^{\scriptscriptstyle AVV}$ 
(c) symmetric notation for $T^{\scriptscriptstyle AAA}$ 
\end{center}

If dimensional regularization is used, 
it is the ambiguity of defining $\gamma_5$ away from four dimensions, rather than the ambiguity of external momentum routes, that becomes the problem in
defining  $T^{\scriptscriptstyle AVV}$, $T^{\scriptscriptstyle AAA}$, and related chiral amplitudes. The definition of massless fermion chiral amplitudes will be an essential part of our discussion of the physical significance of anomaly poles
and, for this purpose, it will be very important that we start with the fermion
mass $m \neq 0$ and that, in effect, we use Pauli-Villars regularization. Indeed, as we emphasize in the main body of this paper, all of the ``massless'' QUD diagrams that we consider are defined via zero mass limit(s) of massive diagrams.

\subhead{A.2 Invariant Amplitudes and Ward Identities.}

$T^{\scriptscriptstyle AVV}$ 
can be decomposed into invariant amplitudes by writing 
$$
\eqalign{T^{\scriptscriptstyle AVV}_{\mu \alpha \beta}(k_1,k_2) ~&= ~ A_1~
{\hbox{\large $\epsilon$}}_{\sigma\alpha\beta\mu}~ k_1^{\sigma}  ~+~ A_2~ 
{\hbox{\large $\epsilon$}}_{\sigma\alpha\beta\mu} ~k_2^{\sigma} 
~+~A_3~
{\hbox{\large $\epsilon$}}_{\delta \sigma\alpha\mu}~ 
k_{1\beta}k_1^{\delta} k_2^{\sigma}  \cr
~~~& +~A_4~  {\hbox{\large $\epsilon$}}_{\delta \sigma\alpha\mu}
~ k_{2\beta}k_1^{\delta}
k_2^{\sigma}~+~A_5~  {\hbox{\large $\epsilon$}}_{\delta \sigma\beta\mu}
~k_{1\alpha}k_1^{\delta}
k_2^{\sigma}~+~A_6~ {\hbox{\large $\epsilon$}}_{\delta \sigma\beta\mu} 
~ k_{2\alpha}k_1^{\delta}
k_2^{\sigma} }
\auto\label{inde}
$$
The Bose symmetry implies that
$$
\eqalign{A_1(k_1,k_2)&=-A_2(k_2,k_1),\cr  
A_3(k_1,k_2)&=-A_6(k_2,k_1), \cr
A_4(k_1,k_2)&=-A_5(k_2,k_1)}
\auto\label{bsym}
$$
The Ward identities corresponding to conservation of the vector currents are 
$$
k_1^{\alpha}~\Gamma_{\mu \alpha \beta}~=0 ~,~~~
k_2^{\beta}~\Gamma_{\mu \alpha \beta}~=0
\auto \label{vwi}
$$
which will be satisfied, respectively, if 
$$
A_2~=~k_1^2~A_5 ~+~k_1\cdot k_2 ~A_6
\auto\label{vwi1}
$$
and 
$$
A_1~=~k_2^2~A_4 ~+~k_1\cdot k_2 ~A_3
\auto\label{vwi2}
$$

In our reggeon diagram analysis, we are particularly interested in the ``Ward identity zeroes'' 
that follow from (\ref{vwi}). In general, if a vector current amplitude  
$\VEV{A_{\mu}(k)~...~}$ satisfies a Ward identity
$$
k^{\mu}~\VEV{A_{\mu}(k)~...~}~=~0
\auto\label{wd2}
$$
the amplitude necessarily vanishes at $k_{\mu}=0$. To see this, we simply
differentiate the Ward identity (treating each component of $k$ as 
independent) to obtain 
$$
\VEV{A_{\mu}~...~}~+~\Biggl[{\partial \VEV{A_{\nu}~...~}\over 
\partial k_{\mu}}\Biggr]_{k=0}~k^{\nu}~=0 \auto\label{wd1}
$$
$$
=>\VEV{A_{\mu}~...~}~~\centerunder{$\to$} {\raisebox{-5mm} 
{$k_{\mu}\to  0$}}0~~~~~~~if~~ ~\Biggl[{\partial \VEV{A_{\nu}~...~}\over 
\partial k_{\mu}}\Biggr]_{k=0} ~
~\st{\to}~\infty
\auto\label{wd3}
$$
implying that, in general, the amplitude should vanish at zero momentum.

When the surface contribution to (\ref{tamp}) is
chosen to ensure that the vector Ward identities (\ref{vwi}) are satisfied
the result (as we discuss next) is a constant term in 
$A_1$ and $A_2$, giving
$$
T^{\scriptscriptstyle AVV}_{\mu \alpha \beta}(k_1,k_2) ~= ~ {1\over 4 \pi^2}~
{\hbox{\large $\epsilon$}}_{\sigma\mu\alpha\beta}~ k_1^{\sigma}  ~-~ 
{1\over 4 \pi^2}~ 
{\hbox{\large $\epsilon$}}_{\sigma\mu\alpha\beta} ~k_2^{\sigma} ~~+~~\cdots
\auto\label{uvco}
$$ 
For momenta where the other $A_i$ are non-singular, only $A_1$ and $A_2$ can
contribute an axial current divergence. The two terms (\ref{uvco}) produce the familiar ``anomaly'' divergence, or anomalous Ward identity,
$$
(k_1 + k_2)^{\mu}~T_{\mu \alpha \beta}~=~
{1\over 2 {\pi}^2 }~{\hbox{\Large $\epsilon$}}_{\sigma\delta\alpha\beta} 
~k_1^{\sigma} k_2^{\delta} + O(m)
\auto\label{awi}
$$

The vector Ward identities (\ref{vwi}) are separately satisfied
by a single term of the form 
$$
 A(k_1,k_2)~
{\hbox{\large $\epsilon$}}_{\delta \sigma\alpha\beta}~ k_1^\delta 
k_2^{\sigma}~(k_1 + k_2 )_{\mu}  
\auto\label{exa}
$$
which also, potentially, contributes to the axial current divergence. Indeed, 
if we introduce an ``anomaly pole'' into $A(k_1,k_2)$ we can obtain a single term 
$$
\frac{1}{2\pi^2 (k_1+ k_2)^2}
{\hbox{\large $\epsilon$}}_{\delta \sigma\alpha\beta}~ k_1^\delta 
k_2^{\sigma}~(k_1 + k_2 )_{\mu}  
\auto\label{exa1}
$$
which directly reproduces the anomaly divergence while, simultaneously, satisfying the vector Ward identities. This will become increasingly significant in the following.

A priori, it might seem that the general term (\ref{exa})
could be added to (\ref{inde}). However, because of the identity
$$ 
\eqalign{~~~{\hbox{\Large $\epsilon$}}_{\delta\sigma\alpha\beta} 
 k_1^{\delta} k_2^{\sigma} [k_1 + k_2]_{\mu} ~=~&
- ({\hbox{\Large $\epsilon$}}_{\delta \sigma\alpha\mu}k_{1\beta}
 -{\hbox{\Large $\epsilon$}}_{\delta \sigma\beta\mu}k_{2\alpha} 
 -{\hbox{\Large $\epsilon$}}_{\delta \sigma\beta\mu}k_{1\alpha}
 +{\hbox{\Large $\epsilon$}}_{\delta \sigma\beta\mu} 
k_{2\beta}) k_1^{\delta}k_2^{\sigma}\cr
 & ~+~{\hbox{\Large $\epsilon$}}_{\sigma\alpha\beta\mu}k_2^{\sigma}~
 (k_1\cdot k_2 - k_1^2)
~-~{\hbox{\Large $\epsilon$}}_{\sigma\alpha\beta\mu}k_1^{\sigma}~ 
(k_1\cdot k_2 -k_2^2)}
\auto\label{epid}
$$
(\ref{exa}) can be re-expressed in the form (\ref{inde}). 
We will appeal to this identity at several points in our discussion.

\subhead{A.3 Surface Term Contributions}

To aid our discussion, we briefly describe how the large momentum surface
contribution to (\ref{tamp}) depends on the routing of external momenta through
the diagram and how this can be fixed by 
requiring the vector Ward identities be satisfied. 
(\ref{uvco}) can be derived from (\ref{tamp}) as follows.

A shift of the
integration variable $p$ by $\delta p = k_2 - k_1$
produces a surface contribution (that is independent of $m$)
$$
\eqalign{\delta \Gamma_{\mu \alpha \beta}(k_1,k_2) ~
&=~(k_1 - k_2)^{\lambda}\int \frac{d^4p}{(2\pi)^4)} 
\frac{\partial}{\partial p_{\lambda}}
 \frac{Tr \{ \gamma_5
\gamma_{\mu} (\st{p} + \st{k}_2)  \gamma_{\alpha}(\st{p} - \st{k}_1 + \st{k}_2 ) 
\gamma_{\beta} (\st{p} - \st{k}_1 ) \}} 
{(p+ k_2 )^2 (p - k_1 + k_2)^2  
 (p - k_1)^2 } \cr
&=~ (k_1 - k_2)^{\lambda}~ \frac{2i\pi^2}{(2\pi)^4}~ 
\centerunder{$lim$}{\raisebox{-3mm}{${\scriptstyle p \to \infty}$}}
~\frac{p_{\lambda}}{p^4}~ Tr\{\gamma_5 \gamma_{\mu}
\gamma_{\rho}\gamma_{\alpha}\gamma_{\delta}\gamma_{\beta}\gamma_{\tau}\}
~p^{\rho}p^{\delta}p^{\tau}}
\auto\label{sti}
$$
If we then write $p_{\lambda}p_{\rho}/p^2 = g_{\lambda \rho}$ we obtain
$$
\frac{1}{8\pi^2}~\epsilon_{\mu\alpha\beta\delta}~(k_1 - k_2)^{\delta}
\auto\label{sti1}
$$
Adding the contribution from  $\Gamma_{\mu \beta \alpha}(k_2,k_1)$ gives 
(\ref{uvco}), with the remaining part of the amplitude defined via the 
(more natural) momentum routing shown in Fig.5.2(b).

A further role is played by surface terms in using (\ref{sti1}) to obtain 
the vector Ward Identities. Evaluating $\Gamma_{\mu \alpha \beta}(k_1,k_2)$
with the momentum routing of Fig.~5.2(b) and using
$$
\st{k}_1 ~=~ (\st{p} - m) ~-~ (\st{p} - \st{k}_1 - m)
\auto\label{ag1}
$$
we obtain 
$$ 
\eqalign{k_1^{\alpha}~\Gamma_{\mu \alpha \beta}(k_1,k_2)~=~
{1 \over (2 \pi)^4 }& \int d^4p~  Tr \{ \gamma_5
\gamma_{\mu} \frac{1}{(\st{p} + \st{k}_1 -m  )}  
\gamma_{\beta} \frac{1}{(\st{p} -\st{k}_2  -m)} \} \cr
& ~-~
{1 \over (2 \pi)^4 } \int d^4p~ Tr \{ \gamma_5
\gamma_{\mu} \frac{1}{(\st{p} -m )} 
\gamma_{\beta} \frac{1}{(\st{p} -\st{k}_2  -m )} \}}
\auto\label{tampr}
$$
When $k_1^{\alpha}~\Gamma_{\mu \beta \alpha}(k_2,k_1)$ is added the result
is a sum of two terms, each of which gives a surface contribution.
The first contribution is
$$ 
\eqalign{{1 \over (2 \pi)^4 } \int d^4p~\biggl(  Tr \{ \gamma_5
\gamma_{\mu} &\frac{1}{(\st{p} + \st{k}_1  -m )}  
\gamma_{\beta} \frac{1}{(\st{p} -\st{k}_2  -m )} \} \cr
& -~  Tr \{ \gamma_5
\gamma_{\mu} \frac{1}{(\st{p} - \st{k}_1  -m )}  
\gamma_{\beta} \frac{1}{(\st{p} +\st{k}_2  -m )} \} \biggr)}
\auto\label{tamp0}
$$
which, since the two integrands differ by an integration shift of
$p \to p -k_1 + k_2$, gives
$$
\eqalign{&(k_1 - k_2)^{\lambda}\int d^4p \frac{\partial}{\partial p_{\lambda}}
\biggl({ Tr \{ \gamma_5
\gamma_{\mu} (\st{p} + \st{k}_1 )  
\gamma_{\beta} (\st{p} -\st{k}_2 ) \} 
\over (p + k_1)^2 (p - k_2)^2 } \biggr)\cr
&= (k_1 - k_2)^{\lambda}~\frac{2i\pi^2}{(2\pi^4)} 
\centerunder{$lim$}{\raisebox{-3mm}{${\scriptstyle p \to \infty}$}}
~\biggl[\frac{p_{\lambda}p^{\rho}}{p^2}\biggr]~\biggl( Tr\{\gamma_5 \gamma_{\mu}
\gamma_{\rho}\gamma_{\beta}\gamma_{\tau}\}
~(-k_1)^{\tau} + Tr\{\gamma_5 \gamma_{\mu}\gamma_{\tau}\gamma_{\beta}\gamma_{\rho}\}
(k_2)^{\tau}\biggr) \cr
& = ~\frac{1}{8\pi^2} \epsilon_{\mu\beta\delta\sigma}~k_1^{\delta}k_2 ^{\sigma}}
\auto\label{st}
$$
Combining this with the second surface contribution and adding the contributions
from terms of the form of (\ref{sti1}) to the full integral,
we obtain the first Ward identity of (\ref{vwi}). 

The Bose symmetry is essential for the appearance of surface integrals.
Note also that, in effect, the kinematic form of the surface contribution to (\ref{tamp0}) is obtained by simply keeping the external momentum contributions to propagator numerators, i.e.
$$
Tr \{ \gamma_5
\gamma_{\mu} \st{k}_1   
\gamma_{\beta} \st{k}_2  \} ~~\to~~
\epsilon_{\mu\beta\delta\sigma}~k_1^{\delta}k_2 ^{\sigma}
\auto\label{numc}
$$

If Pauli-Villars regularization is used then the integral defined by subtracting an unphysical finite mass fermion loop is convergent and so has no surface term
contributions. Therefore, all integration variable shifts go through straightforwardly and both (\ref{sti}) and (\ref{st}) are zero. Consequently, 
both the axial and vector current are conserved.
The anomaly appears (in the axial current divergence) as a remnant when the infinite mass limit is taken for the unphysical fermion. As we will discuss further, a contribution of the form (\ref{st}), which cancels the anomaly in the vector current divergence, can also be produced by the unphysical fermion loop.

\subhead{A.4 The Anomaly Pole Via the Ward Identities}
 
There is a close
relationship between the ultra-violet anomaly contribution (\ref{uvco}) that appears in $A_1$ and $A_2$ and an  ``anomaly pole'' that appears in $A_3$ and $A_6$ and plays an essential role in the satisfaction of the vector Ward Identities. 
That such a pole might be present 
can be seen immediately by noting that if $k_1^2=0$ 
and $A_5$ is not singular at this point, then
the Ward identity (\ref{vwi1}) reduces to the very simple
form  
$$
A_6~=~ \frac{A_2}{k_1.k_2} ~=~{2 \over (q^2 - k_2^2)}~ A_2
\auto\label{vwi3}
$$
Inserting the contribution obtained from (\ref{uvco}) then gives
$$
 A_6~=~-~ {1 \over 2\pi^2 (q^2 -k_2^2)}~+~...  
\auto\label{vwi4}
$$
suggesting, at first sight, that there should be a pole at 
$q^2 = k_2^2$, independently of the value of $m^2$ !

The second Ward identity suggests (only a little less straightforwardly) that  the same pole should also be present in $A_3$. If we set both $k_1^2=0$ and $k_2^2=0$, the situation is even simpler. 
The Ward identities (\ref{vwi1}) and (\ref{vwi2}) now reduce to
$$
A_3~=~ {2 \over q^2}~A_1 ~=~ \frac{A_1}{k_1.k_2}~, ~~~~ A_6~=~ {2 \over q^2}~ A_2  
~=~\frac{A_2}{k_1.k_2}
\auto\label{vwi5}
$$
Inserting, again, the contributions from (\ref{uvco}) gives
$$
A_3~=~ {1 \over 2\pi^2 q^2}~+ ~...~~, ~~~~ A_6~=~-~ {1 \over 2\pi^2 q^2}~+~...  
\auto\label{vwi6}
$$
suggesting even more strongly that, in this special kinematic situation, there should be a pole at $q^2=0$ (for arbitrary $m^2$ !)

In fact, we will find that the pole (\ref{vwi6}), and the pole (\ref{vwi4}), cancel in physical amplitudes in almost all kinematic configurations and are never present when $m^2\neq 0$. Although we will show that, when $k_1^2=0$, a singularity at $q^2 = k_2^2$ does have a very real physical significance. Our results will be consistent with the results of \cite{bfsy}, where it is shown
that if dispersion relation representations
for all the invariant functions are inserted into the Ward identities, the vector
identities are satisfied only if the pole (\ref{vwi6}) is indeed 
present at $q^2=0$ when $k_1^2 = k_2^2 =0$, but only when $m=0$.  
For our purposes, however, a much larger issue is whether there are circumstances in which this pole can be interpreted as a Goldstone boson particle associated with
the spontaneous breaking of a chiral symmetry generated by an axial-vector current.

Note that we can equally well carry through the above arguments if  
$k_1^2, k_2^2 \neq 0$,
provided that $q^2 >> k_1^2, k_2^2$. We then obtain the 
asymptotic result
$$
T^{\scriptscriptstyle AVV}_{\mu \alpha \beta}(k_1,k_2) ~~
\centerunder{$\rightarrow$}{\raisebox{-5mm}{$ q^2 \to \infty$}}
~~ \frac{1}{4\pi^2}(
{\hbox{\large $\epsilon$}}_{\sigma\alpha\beta\mu} k_1^{\sigma} -  
{\hbox{\large $\epsilon$}}_{\sigma\alpha\beta\mu} k_2^{\sigma} )
+\frac{1}{4\pi^2 k_1.k_2}(
{\hbox{\large $\epsilon$}}_{\delta \sigma\alpha\mu} 
k_{1\beta}k_1^{\delta} k_2^{\sigma}- {\hbox{\large $\epsilon$}}_{\delta \sigma\beta\mu} 
 k_{2\alpha}k_1^{\delta}k_2^{\sigma}) 
\auto\label{inde0}
$$
The two terms in (\ref{inde0})
are sufficient, in themselves, to both satisfy the vector Ward identities 
and give the anomaly in the axial divergence. The ``anomaly pole'' term reproduces the
surface contribution to the vector Ward identity that we obtained above. Note that,
as $q^2 \to \infty$ (with $k_1^1, k_2^2$ and $m^2$ finite),
any finite momentum region of (\ref{tamp}) contributes behavior of the form
$$
A_1 ~~\centerunder{$\sim$}{\raisebox{-5mm}{$ q^2 \to \infty$}}~~O\bigl(\frac{1}{q^2}\bigr)
~~, ~~~~
A_3 ~~\centerunder{$\sim$}{\raisebox{-5mm}{$ q^2 \to \infty$}}~~O\bigl(\frac{1}{q^4}\bigr)
\auto\label{fq}
$$
and so both terms in (\ref{inde0}) must originate from an internal
region involving large momentum components. 

If we assume that $A_4$ and $A_5$ contributions can consistently be neglected 
when $k_1^2$ and $k_2^2$ are small, as the Ward identities (\ref {vwi}) imply, (and as is more generally the case for the reggeon vertices discussed in the text), then (\ref{epid}) can be used to simplify (\ref{inde0}) to the simple form (\ref{exa1}), i.e.
$$
T^{\scriptscriptstyle AVV}_{\mu \alpha \beta}(k_1,k_2) ~~
\centerunder{$\rightarrow$}{\raisebox{-5mm}{$ q^2 \to \infty$}}
~~ ~~q_{\mu}~\frac{
{\hbox{\large $\epsilon$}}_{\delta \sigma\alpha\beta}~ k_1^\delta 
k_2^{\sigma} }{2\pi^2 q^2}
\auto\label{indes}
$$
This expression is particularly suggestive for our purposes because it has the appropriate factorization property for the
pole to be interpreted as a particle pole in the $q^2$ channel.
At this point, however, it is an asymptotic result for large $q^2$ that clearly does not imply the presence of a pole at $q^2=0$. Indeed, it would surely appear paradoxical if an infra-red particle pole,
at $q^2=0$, is generated in the infinite momentum part of the
integral - independently of the existence of massless particles
in the amplitude. Nevertheless, we might expect some complication from the fact 
that we are subtracting the 
contribution of infinitely massive particles. In fact, as we will see in the next sub-section that there is a direct connection between a large (light-cone) 
momentum generated $1/q^2$ and an infra-red pole. 

\subhead{A.5 The Anomaly Pole in $T^{\scriptscriptstyle AAA}$}

The analysis of $T^{\scriptscriptstyle AAA}_{\alpha \beta\gamma}(k_1,k_2,k_3)$
in \cite{cg} is very important for our discussion, both in this subsection and later. We begin by noting that, in this analysis, it is assumed that
{\it $m^2$ has been set to zero,} and that the amplitude 
is completely symmetrized with respect to simultaneous permutations of
$(k_1,k_2,k_3)$ and $(\alpha, \beta, \gamma)$. In this case, if we set
$$
k_1^2 = k_2^2 = k_3^2 = Q^2
\auto\label{lin}
$$
the symmetry implies that we can write $T^{\scriptscriptstyle AAA}$ as a sum of terms of the form (\ref{exa}), i.e.
$$
T^{\scriptscriptstyle AAA}_{\alpha \beta\gamma}(k_1,k_2,k_3)
= F(Q^2)~({\hbox{\large $\epsilon$}}_{\alpha\beta\delta \sigma} 
~ k_1^{\delta}k_2^{\sigma} k_{3\gamma} + 
{\hbox{\large $\epsilon$}}_{\beta\gamma\delta \sigma} 
~ k_2^{\delta}k_3^{\sigma}
k_{1\alpha} +
{\hbox{\large $\epsilon$}}_{\gamma\alpha\delta \sigma} 
~ k_3^{\delta}k_1^{\sigma}
k_{2\beta})
\auto\label{lin1}
$$
The anomaly divergence then becomes
$$
k_3^{\gamma} ~T^{\scriptscriptstyle AAA}_{\alpha \beta\gamma}
~=~ Q^2 F(Q^2) ~
{\hbox{\large $\epsilon$}}_{\alpha\beta\delta \sigma} 
~ k_1^{\delta}k_2^{\sigma}
~=~\frac{1}{2\pi^2} ~ {\hbox{\large $\epsilon$}}_{\alpha\beta\delta \sigma} 
~ k_1^{\delta}k_2^{\sigma}
\auto\label{lin2}
$$
implying that 
$$
F(Q^2)~=~ \frac{1}{2\pi^2 Q^2} ~~~~~~~~\hbox{\large \it for~all $~Q^2$}
\auto\label{lin3}
$$

Therefore, in this very special kinematic situation
$T^{\scriptscriptstyle AAA}$ can be written as a sum of three terms, each of which has the form (\ref{indes}) and so contains an anomaly pole in one channel and has zero divergence in the other two channels. Moreover this sum of anomaly pole amplitudes gives the complete amplitude.  
This will be very important later when we discuss chiral amplitudes
even though,
because of the difficulty of combining $\gamma_5$ manipulations with the anomaly (that we will discuss later) there is  
not an immediate connection between $T^{\scriptscriptstyle AAA}$ and $T^{\scriptscriptstyle AVV}$.

That there is no sign of the particle thresholds in $F(Q^2)$ is not surprising since there is no invariant to set the scale for logarithms. Indeed, without the anomaly the amplitude would have been zero. Instead, it appears that
the anomaly pole, that we located at large $q^2$ and small $k_1^2, k_2^2$, in the previous 
sub-section, is present for all $Q^2$, when (\ref{lin}) is satisfied, down to $Q^2=0$. So, is it generated at large or small momentum? In fact, as we will see 
from the explicit formulae that we introduce in the next sub-section, the situation is actually quite subtle. Our analysis 
will explain how the thresholds have disappeared and will amplify the argument in \cite{cg} that the pole in $F(Q^2)$ 
is due to massless particles contributing via the triangle singularity.

At first sight, (\ref{lin3}) does not seem to necessitate a physical singularity in $T^{\scriptscriptstyle AAA}$. Rather,
from (\ref{lin1}), if $k_1, k_2$ and $k_3$ are all spacelike 
$$
k_1 \sim k_2 \sim k_3 \sim Q ~\to 0 ~~\implies~~
T^{\scriptscriptstyle AAA} ~\centerunder{$\sim$}{\raisebox{-4mm}{${\scriptstyle Q^2 \to 0}$}} ~~Q
\auto\label{lin4}
$$ 
However, as is observed in \cite{cg}, this is not the case if a {\it finite} light-like vector, outside of the existing momentum plane, is added to 
$k_1$, say. For example, if we define basis vectors
$$
e_1 ~=~(0,1,0.0), ~~~e_2 ~=~(0,0,1,0),~~~
e_{\pm}~=~(1,0,0,\pm 1)/ \sqrt{2},
\auto\label{bvs}
$$
and write
$$
k_1 = Qe_1 + e_+,~~~ k_2=Q(-e_1 +\sqrt{3})/2, ~~~k_3= - Q(e_1 +\sqrt{3})/2
-e_+
\auto\label{+lc}
$$
we obtain
$$
T^{\scriptscriptstyle AAA} ~\centerunder{$\sim$}{\raisebox{-4mm}{${\scriptstyle Q^2 \to 0}$}} ~~
\frac{1}{Q}
\auto\label{lin5}
$$ 
and so a physical singularity must be present.
(This is closely related to our argument, in the following, that in an infinite momentum frame a Goldstone boson can appear as an anomaly pole.) We will describe later how the triangle diagram internal momentum configuration that gives the anomaly pole is identified in \cite{cg}. 

Next, we describe some explicit formulae for the invariant functions $A_i$
appearing in
$T^{\scriptscriptstyle AVV}$ that have been derived in the literature. These formulae will provide us with considerable insight
into how an anomaly pole can appear as both a large and a small momentum phenomenon.

\subhead{A.6 Explicit Formulae} 

An analytic expression for the full amplitude
(\ref{tamp}) can be found in \cite{cor}. For our purposes we can use simpler expressions that are valid in a special kinematic configuration. Motivated by our 
Ward Identity discussion of the anomaly pole, we
set one vector current invariant $k_1^2$ to zero. This also corresponds to an anomaly vertex residue of the anomalous wee gluon divergence discussed in the paper. We use the following set of formulae, given in \cite{ach}, that hold when $k_1^2 = 0$, with $k_2^2,~ q^2 < 0 $ and $ m^2 >0$.
$$
\eqalign{A_6~&=~-A_3~= ~- ~{1 \over 2\pi^2}{1 \over k^2_2 -q^2} \biggl(
{k_2^2 \over k_2^2 -q^2 } L_1 ~-~ {m^2 \over k_2^2 -q^2 } L_2 ~- ~1\biggr) \cr
A_4~&= ~ {1 \over 2\pi^2}{1 \over k^2_2 -q^2 } ~ L_1 \cr
A_2~&= ~ {1 \over 4\pi^2}\biggl(
{k_2^2 \over k_2^2 -q^2} L_1 ~-~ {m^2 \over k_2^2 -q^2 } L_2 ~- ~1\biggr)\cr
A_1~&= ~ {1 \over 4\pi^2}\biggl(
{k_2^2 \over k_2^2 -q^2 } L_1 ~+~ {m^2 \over k_2^2 -q^2 } L_2 ~+ ~1\biggr)\cr
A_5 &= -A_4 - {3 \over \pi^2}k_2^2 {d \over d k_2^2}\biggl( {1 \over   
k_2^2 - q^2 } L_1 \biggr) + 
{3 \over 2\pi^2}k_2^4 \biggl({d \over d k_2^2}\biggr)^2\biggl( {1 \over   
k_2^2 - q^2 } L_1 \biggr) \cr
& ~~~~ + {3 \over 4\pi^2}k_2^2 {d \over d k_2^2}\biggl( {1 \over   
k_2^2 -q^2  } L_2 \biggr) +
{1 \over 2\pi^2}m^2 k_2^2 \biggl({d \over d k_2^2}\biggr)^2
\biggl( {1 \over k_2^2 -q^2 } L_2 \biggr)}
\auto\label{k2q}
$$
where
$$
\eqalign{ &~~~~L_1 ~=~ -~ \rho~ \ln{ \rho +1 \over \rho -1} ~+~
 \beta ~\ln{ \beta +1 \over \beta -1} \cr
&~~~~ L_2 ~=~ - ~\ln^2{ \rho +1 \over \rho -1}~ + ~ \ln^2{ \beta +1 \over \beta -1} \cr
&~~~~ \rho^2~=~ 1 ~- ~ 4m^2/q^2 ~,~~ ~~ \beta^2~=~ 1 ~- ~ 4m^2/k_2^2 }
\auto\label{L1L2}
$$

The simplified Ward identity (\ref{vwi3}) is explicitly satisfied and both terms in (\ref{inde0}) are directly present in the appropriate  
invariant functions and dominate at large $q^2$, as anticipated. At first sight, there is also a simple ``anomaly pole''
at $q^2=k_2^2$, in both $A_6$ and $A_3$, as predicted by 
(\ref{vwi4}). However, we will show in the next subsection, by taking various limits of (\ref{k2q}), that in most kinematic circumstances this pole is actually canceled
in the complete functions.

First, it will be helpful to identify the nature of the various contributions to the $A_i$ in terms of singularity content. Even without formulating an appropriate dispersion relation, we can obviously write  
$$
A_i =~ A^i_{q^2} ~+~A^i_{k_2^2} ~+~ A^i_{\infty} ~~~~~~~i~=~1,2,3,6
\auto\label{sep}
$$
where, for example,
$$
\eqalign{
A^1_{q^2}&=~{1 \over 4\pi^2 } \biggl(
{k_2^2 \over k_2^2 -q^2 } [~ \rho~ \ln{ \rho +1 \over \rho -1}] 
 ~-~ {m^2 \over k_2^2 -q^2 } [~\ln^2{ \rho +1 \over \rho -1}] \biggr)\cr
A^1_{k_2^2}&=~{1 \over 4\pi^2} \biggl(
{k_2^2 \over k_2^2 -q^2 } [~ \beta~ \ln{ \beta +1 \over \beta -1}] 
 ~-~ {m^2 \over k_2^2 -q^2 } [ ~\ln^2{ \beta +1 \over \beta -1}] \biggr)\cr
A^1_{\infty}&=~{1 \over 4\pi^2} }
\auto\label{sep1}
$$
$A^1_{q^2}$ can be identified as the contribution of the $q^2$ threshold, $A^1_{k_2^2}$ is the contribution of the $k_2^2$ threshold and $A^1_{\infty}$ is the anomaly contribution from infinite momentum.  
The separation (\ref{sep}) in each of $A_2$, $A_3$ and 
$A_6$ is analagous. 

In $A_3$ and $A_6$ an anomaly pole is present in what appear to be  infinite momentum contributions even though, as we will see, in general the double poles at $q^2=k_2^2$  in the threshold contributions to $A_3$ and $A_6$
combine to cancel the infinite momentum pole contributions. Clearly, we would like to understand whether such a pole can actually be identified as an infinite momentum contribution. 

\subhead{A.7 Limits}

We begin with an ultra-violet limit which shows that $A^1_{\infty}$
can be consistently identified as originating from the infinite mass limit of an unphysical fermion loop. We consider the $m^2 \to \infty$ limit of
$A^1_{q^2} + A^1_{k_2^2}$. In this limit $\rho, \beta \to \infty$ and since 
$$
\ln\frac{\rho +1}{\rho -1} ~\to~ \ln (1 + \frac{2}{\rho} + ...)~~
\to ~~ \frac{2}{\rho} 
\auto\label{qt0}
$$
we obtain
$$ 
\eqalign{A^1_{q^2} + A^1_{k_2^2} ~~\centerunder{$\longrightarrow$}{
\raisebox{-5mm}{$m^2 \to ~\infty$}} ~~ 
 &{m^2 \over 4 \pi^2 (k_2^2 -q^2 )} \biggl[\ln^2{ \rho +1 \over \rho -1}~- ~\ln^2{ \beta +1 \over \beta -1}\biggr]\cr  
~& = {m^2 \over 4 \pi^2 (k_2^2 -q^2 )} \biggl[\frac{-k_2^2}{m^2} - ~
\frac{-q^2}{m^2} \biggr] ~=~-~\frac{1}{4\pi^2}}
\auto\label{laM}
$$

Interpreting $A^1_{q^2} + A^1_{k_2^2}$ as the contribution of the unsubtracted amplitude, we see that  
$A^1_{\infty}$ acts as a Pauli-Villars (opposite sign fermion loop) subtraction which ensures that $A^1 \to 0$ when $m^2 \to ~\infty$. Apparently,
if we interpret
$$ 
A^6_{q^2} + A^6_{k_2^2} ~~\centerunder{$\longrightarrow$}{
\raisebox{-5mm}{$m^2 \to ~\infty$}} ~~ 
-~\frac{1}{2\pi^2 (k_2^2 - q^2)}
\auto\label{laM1}
$$
as the contribution of the unsubtracted $A_6$ amplitude, then
the anomaly pole term $A^6_{\infty}$ plays an identical role in $A^6$ to that played by the anomaly in $A^1$. In particular, the pole at $q^2=k_2^2$, that would otherwise be present at large $m^2$, is canceled.
We conclude that it should be possible to interpret the infinite momentum pole term as having it's origin in the Pauli-Villars unphysical fermion loop subtraction. This would be natural, since it is the 
addition of the unphysical loop that gives an amplitude which straightforwardly
satisfies the vector current Ward identity and results in the simple relation between $A_1$ and $A_6$ which (at $k_1^2 =0$) directly relates the anomaly pole to the anomaly. We will return to
this discussion once we have described the connection between the anomaly pole and
the triangle Landau singularity.

We consider next some relatively simple infra-red limits. First we consider, directly, the 
$q^2 \to 0$ limit. Again using (\ref{qt0}), 
$$
A_6 ~\to~ \frac{1}{2\pi^2k_2^2}\biggl(-2 + \beta\ln\frac{\beta +1}{\beta -1}
- \frac{m^2}{k_2^2}\ln^2\frac{\beta+1}{\beta-1} -1\biggr)
\auto\label{qt01}
$$
and so there is clearly no pole at $q^2=0$. If we now take the further limit $k_2^2 \to 0$ we obtain
$$
\eqalign{A_6 ~&\to~ 
\frac{1}{2\pi^2k_2^2}\biggl(-\frac{m^2}{k_2^2}\biggl[\frac{2}{\beta}\biggr]^2 -1 
\biggr) ~+~ O(\frac{1}{m^2}) \cr
~&=~ \frac{1}{2\pi^2k_2^2}\biggl(1~-~1\biggr)~+~O(\frac{1}{m^2})~=~O(\frac{1}{m^2})}
\auto\label{qt02}
$$
and again there is no pole. Taking all invariants to zero, with $m^2$ finite,
is effectively the same as the $m^2 \to \infty$ limit and so 
produces the same cancellation (involving the infinite momentum pole). Very importantly, however, we see that when all momentum invariants are zero the 
amplitude is singular at $m^2 = 0$, as the dimension of the amplitude implies it must be. If we take $m^2 \to 0$ before taking $k_2^2 \to 0$,
there is a logarithmic divergence and no limit is obtained.
Clearly, the combination of the zero mass and zero momentum limits is ambiguous and depends on the order of the limits.

If we take $ m^2 \to 0$ directly in (\ref{k2q}) we obtain 
$$
A_6~=~- {1 \over 2{\pi}^2 }~{1 \over k_2^2 -q^2}
 \biggl({k_2^2 \over k_2^2 - q^2}~\ln{k_2^2
\over q^2} ~-~1 \biggr) 
\auto\label{k1m0}
$$
The large momentum pole term survives straightforwardly, while the remainder gives a non-zero limit only because of the singularity of the threshold terms.
Now, it is the limit $q^2 \to 0$ that gives a logarithmic divergence. If we first take the limit $k^2 \to 0$, only the pole term survives, giving the whole amplitude i.e.
$$
A_6~=~- {1 \over 2{\pi}^2q^2 }
\auto\label{ksm0}
$$
in agreement with (\ref{vwi6}). Since $m^2=0$, this limit is equivalent to the 
limit of large $q^2$, with finite $k_1^2$ and $k_2^2$. Consequently, the surviving pole is from the large momentum contribution. The
``infra-red'' momentum contributions containing the logarithmic thresholds 
have simply dropped away.  

Note that if the $q^2$ or the $k_2^2$ discontinuity is taken in (\ref{k1m0}), so that the 
logarithm aquires an imaginary part, the infra-red threshold contribution gives a double pole at $q^2 = k^2$ that we will shortly identify with the triangle Landau singularity. 

If we take the limit $k_2^2 \to q^2$
in (\ref{k1m0}), we again obtain an amplitude which contains only the $q^2 =0$ pole, i.e. 
$$
\eqalign{A_6~&=~-
{1 \over 2{\pi}^2 }{1 \over k_2^2 -q^2}
 \biggl({k_2^2 \over k_2^2 - q^2}~\bigl(~ \frac{k_2^2 -q^2}{q^2} ~- ~
\frac{(k_2^2 -q^2)^2}{2q^4}~ +~ ...~\bigr)~-~1 \biggr)\cr 
&=~- {1 \over 4{\pi}^2q^2 }}
\auto\label{k1qk0}
$$
but the residue is halved. 
Although $k_2^2 = q^2$ has a deeper significance that we will come to, this kinematics parallels, in part,
the restriction (\ref{lin}) that leads to the isolation of the anomaly pole in $T^{\scriptscriptstyle AAA}$. The large momentum anomaly pole term is cancelled and the remaining pole term emerges from the ``infra-red'' logarithmic contribution. At this special kinematic point the logarithmic contributions of the thresholds cancel (rather than dropping away), thus allowing the triangle singularity to emerge as a simple pole at 
$q^2=0$ that is,
as we shall see, a direct consequence of the existence of massless fermions. 
Presumably, (\ref{lin3}) results from a very similar phenomenon. 

The Bose symmetry (\ref{bsym}) implies an analagous contribution to (\ref{k1qk0}) would be obtained by taking
$k_2^2 \to 0$ followed by $k_1^2 \to q^2$. In the symmetric limit 
$q^2 = k_1^2 = k_2^2 ~\to 0$ the two contributions should combine to give the full anomaly. If this is the case,
adding the corresponding contributions from $A_3, A_1,$ and $A_2$ 
(and assuming we can neglect $A_4$ and $A_5$, as we will discuss further in 
{\bf A.12}), gives an infra-red result analagous to (\ref{indes}), i.e. 
$$
T^{\scriptscriptstyle AVV}_{\mu \alpha \beta}(k_1,k_2) ~~
\centerunder{$\longrightarrow$}{\raisebox{-5mm}{\small $ q^2 = k_1^2 = k_2^2 = Q^2~\to~ 0$}}
~~ ~~q_{\mu}~\frac{
{\hbox{\large $\epsilon$}}_{\delta \sigma\alpha\beta}~ k_1^\delta 
k_2^{\sigma} }{2\pi^2 q^2}
\auto\label{indes+}
$$
Therefore, in this symmetric infra-red region, $T^{\scriptscriptstyle AVV}$
will be described by one of the three terms contributing to
$T^{\scriptscriptstyle AAA}$ in (\ref{lin1}). Indeed, since there is only one invariant, i.e $Q^2$, and there is no second scale to provide 
a scale for logarithms, it must be that (\ref{indes+}) describes 
$T^{\scriptscriptstyle AVV}$ for all $Q^2$, just as (\ref{lin1}) describes
$T^{\scriptscriptstyle AAA}$ for all $Q^2$.

If, in (\ref{k2q}), the $ k_2^2 \to 0$ limit is taken first, with $m^2 >0$, the result is
$$
A_6 ~= ~ - {1 \over 2 \pi^2} ~{1 \over q^2} \biggl( 1 +
  ~{m^2 \over q^2} ~\ln^2{ \rho +1 \over \rho -1}~\biggr)
\auto\label{k1k20}
$$
If we now take the limit $q^2 \to 0$, we obtain
$$
A_6 ~\to  ~ - {1 \over 2 \pi^2} ~{1 \over q^2} \biggl( 1 +
  ~{m^2 \over q^2} ~\ln^2{ [1 + ({-q^2 \over m^2})^{1\over 2} + \cdots}] 
~\biggr) ~= ~O(\frac{1}{m^2})~~\st{\longrightarrow}~~\infty 
\auto\label{k1k201}
$$
and so, for $m^2 \neq 0$, the anomaly pole is absent. The only finite $q^2$ 
singularity is the threshold at $q^2 = 4m^2$. As in the closely related $m^2 \to \infty$ limit, the infinite momentum pole cancels with a pole in the part of the amplitude that is directly vanishing as $m^2 \to 0$. Again, 
the limit $m^2 \to 0$ is singular, as the dimension of the amplitude implies it must be, if the limit is taken after $q^2 \to 0$.

Finally, we consider the limit $k_2^2 \to q^2$, with $m^2 \neq 0$, 
and again ask whether
the pole at $k_2^2=q^2$ is present. Because this limit is more 
intricate we give a few more details of the calculation.
Since $\beta \to \rho$ when $k_2^2 \to q^2$, we see from (\ref{L1L2}) that we need
derivatives of $L_1$ and $L_2$, i.e.
$$
\eqalign{~~\frac{dL_1}{d\beta}&~=~ 
\ln \frac{\beta + 1}{\beta -1} - \frac{2\beta}{(\beta^2 -1)},
\cr \frac{d^2L_1}{d\beta^2} &~=~ - \frac{4}{(\beta^2 -1)} + \frac{4\beta^2}{(\beta^2-1)^2}
~=~ 
\frac{4}{(\beta^2-1)^2}\cr
\frac{dL_2}{d\beta}&~=~ -\frac{4}{(\beta^2 -1)}\ln \frac{\beta +1}{\beta - 1}, \cr
\frac{d^2L_2}{d\beta^2}&~=~ \frac{8\beta}{(\beta^2-1)^2}\ln\frac{\beta + 1}{\beta -1} +
\frac{8}{(\beta^2-1)^2}}
\auto\label{dL}
$$
Writing $k^2=q^2 + \delta q^2$, we have 
$$
\eqalign{~~~~~~~~
\delta\beta &= (1-\frac{4m^2}{q^2})^{- \frac{1}{2}} ~\frac{2m^2}{q^4}~ \delta q^2
~-~(1-\frac{4m^2}{q^2})^{- \frac{3}{2}}\frac{2m^2(q^2 -3m^2)}{q^8} ~(\delta q^2)^2
\cr
&=-\frac{\beta^2 - 1}{2\beta} ~\frac{\delta q^2}{q^2} ~+~\frac{\beta^2 -1}{8\beta^3}
(3\beta^2 + 1) ~\bigl(\frac{\delta q^2}{q^2}\bigr)^2  \cr
&= ~ \delta_1 \beta ~+~ \delta_2 \beta}
\auto\label{db}
$$
together with
$$
\frac{k_2^2}{k_2^2-q^2}L_1 - \frac{m^2}{k_2^2-q^2}L_2 =
\frac{q^2 + \delta q^2}{\delta q^2}L_1 + \frac{q^2(\beta^2 -1)}{4 \delta q^2}L_2
\auto\label{d1}
$$
and so the $0\bigl((\delta q^2)^{-1}\bigr)$ contribution to $A_6$ is  
$$
\eqalign{& ~~~~~~ - \frac{1}{2\pi^2}\frac{1}{\delta q^2}\biggl(
\frac{q^2 \delta_1 \beta}{\delta q^2}\biggl[\frac{dL_1}{d\beta} + \frac{\beta^2-1}{4}
\frac{dL_2}{d\beta}\biggr] -1 \biggr)\cr
&=~ - \frac{1}{2\pi^2}\frac{1}{\delta q^2}\biggl(
-\frac{\beta^2-1}{2\beta}\biggl[\ln \frac{\beta +1}{\beta -1} - \frac{2\beta}{\beta^2 -1}
+ \frac{\beta^2-1}{4} \frac{-4}{\beta^2 -1}\ln \frac{\beta +1}{\beta -1}\biggr] -1 
\biggr)\cr 
&~~~~~~~=~ 0} 
\auto\label{obe}
$$
showing that, once again, there is no pole. The infinite momentum pole term is explicitly canceled by the combination of the double poles in the threshold terms.

The finite contribution to $A_6$, as $\delta q^2 \to 0$, is  
$$
\eqalign{ A_6~ &= - \frac{1}{2\pi^2}\frac{1}{\delta q^2}\biggl(\bigl[\delta_1 \beta ]
\biggl[
\frac{dL_1}{d\beta}\biggl]~+~
\frac{q^2 \bigl[\delta_2 \beta\bigr]}{\delta q^2}\biggl[\frac{dL_1}{d\beta} 
+ \frac{\beta^2-1}{4}
\frac{dL_2}{d\beta}\biggr]\cr
&~~~~~~~~~~~~~~~+~\frac{\q^2}{\delta q^2}\frac{\bigl[\delta_1 \beta \bigr]^2}{2}
\biggl[\frac{d^2L_1}{d\beta^2} ~+~\frac{\beta^2 -1}{4} \frac{d^2 L_2}{d\beta^2}\bigg]
\biggl) \cr
&=  -  \frac{1}{2\pi^2 q^2}
\biggl( \frac{1}{4}~+~\frac{1}{4\beta^2}~+~\frac{\beta^2-1}{4\beta^2}
\biggl[1 -~ \beta\ln\frac{\beta+1}{\beta-1}\biggr]\biggr) \cr
&=- \frac{1}{4\pi^2}\biggl(\frac{1}{q^2} +  
2\frac{m^2}{q^4}(1-\frac{4m^2}{q^2})^{-\frac{1}{2}}
\ln \frac{(1-\frac{4m^2}{q^2})^{\frac{1}{2}} + 1}
{(1-\frac{4m^2}{q^2})^{\frac{1}{2}} - 1}~\biggr)
}
\auto\label{obe2}
$$
If we now take the limit $q^2 \to 0$, with $m^2 \neq 0$, we obtain
$$
A_6 ~\to  - \frac{1}{4\pi^2}\biggl(\frac{1}{q^2} +  
2\frac{m^2}{q^4}(-\frac{4m^2}{q^2})^{-\frac{1}{2}}~ 2 
(-\frac{4m^2}{q^2})^{-\frac{1}{2}}\biggr)
~+~... ~= ~O(\frac{1}{m^2})
\auto\label{obe3}
$$
i.e. there is no pole. If we first take $m^2 \to 0$ in (\ref{obe2}),
we obtain  
$$
A_6 ~= ~ - \frac{1}{4\pi^2q^2}
\auto\label{obe4}
$$
which is the same result as (\ref{k1qk0}), and so the limits $k_2^2 \to q^2$ and $m^2 \to 0$
commute. We emphasize, again, that it is the pole contribution from the threshold terms that survives. The thresholds have cancelled, 
leaving (\ref{obe4}) as an exact result, valid for all $q^2 \neq 0$, in agreement with (\ref{lin3}). 

We conclude that the anomaly pole emerges in the invariant functions of the physical triangle diagram amplitude only in very special kinematic circumstances. It appears when $m^2=0$ and the external momentum dependence is reduced to a single invariant.
When the anomaly is included in the residue, it provides the complete amplitude. If we take
$k_1^2,~ k_2^2 \to 0$ before $q^2 \to 0$, the pole that appears
(we have suggested) could, perhaps, be interpreted as an infinite mass Pauli-Villars subtraction, generated at large momentum. If we take $k_1^2 = k_2^2 = q^2 $, or $k_1^2 \to 0,~ k_2^2 \to q^2$, the pole appears from behind the thresholds and is generated at small momentum. The first of these last two
limits is involved in producing momentum conservation for wee gluon vertices,
while the second is involved in the generation of infinite-momentum Goldstone boson particle poles. We will discuss later the role of the external momentum factors 
in determining whether the pole appears in the full amplitude. As we will see,
this is where infinite momentum  plays an essential role.

In the literature there exist\cite{hor} explicit formulae for
the imaginary parts of each of the invariant amplitudes $A_i$ when $k_1^2=k_2^2=k^2~$.
The $q^2$ discontinuity of $A_3$, when $m=0$, is given in \cite{bfsy} in the form
$$
\eqalign{ \centerunder{disc}{\raisebox{-4mm}{$~~~~{\scriptsize q^2}$}}
~ A_3(q^2,k^2) ~=~- \frac{2k^2}{q^2}\theta(q^2)& \biggl[ 
\frac{2k^2(q^2-k^2)}{\sqrt{q^2}(q^2-4k^2)^{5/2}}
\ln \frac{q^2 -2k^2 + \sqrt{q^2(q^2 -4k^2)}}{-2k^2}\cr
& + \frac{q^2 + 2 k^2}{2(q^2-4k^2)^2}\biggr]}
\auto\label{dc3}
$$
This formula can be used\cite{bfsy} to show that
$$
\int_0^{\infty} disc A_3(q^2,k^2)f(q^2) dq^2 ~~\centerunder{$\longrightarrow$}{
\raisebox{-4mm}{$k^2 \to 0$}} ~~  f(0)
\auto\label{itd}
$$
It is the isolated (double) pole term that produces this result, with the region  
$q^2 \sim k^2$ dominating. From the above discussion, we conclude that it is the 
infra-red generated anomaly pole that is involved.

Note that when $q^2=k^2$ the logarithmic threshold term vanishes, in agreement with the above discussion. Also, when $q^2 \to \infty$ it is the isolated pole term that dominates. For reasons that we discuss in the next subsection, when 
$k_1^2 = k_2^2$ it is the limit
$q^2 \to 4k^2$ that should be compared with the limit $q^2 \to k_2^2$ that we have 
discussed for $k_1^2 =0$. 

\subhead{A.8 The Anomaly Pole and the Triangle Landau Singularity}

In this subsection we will provide considerable insight into the above discussion
by directly associating the anomaly pole with the 
triangle Landau singularity. This association will be more general than the argument in \cite{cg} that the discontinuity of the pole at $q^2=k_1^2=k_2^2=0$ is given by the triangle diagram formula. 

The triangle singularity occurs when the 
external momenta are such that all three internal propagators in 
Fig.~A.2 are on-shell and produce poles that trap the integration contour 
at some point in the integration region. The equation for the  singularity surface is obtained by solving the Landau equations
consisting of the three mass-shell conditions
$$
p_i^2~=~m^2~, ~~~~~i=1,2,3
\auto\label{trsm}
$$
together with (what is essentially) a condition that the contour
be trapped, i.e.
$$
\Sigma_i~\alpha_i~p_i ~=~0
\auto\label{trl}
$$
for some set of $\alpha_i$. 
\begin{center}  
\epsfxsize=2in
\epsffile{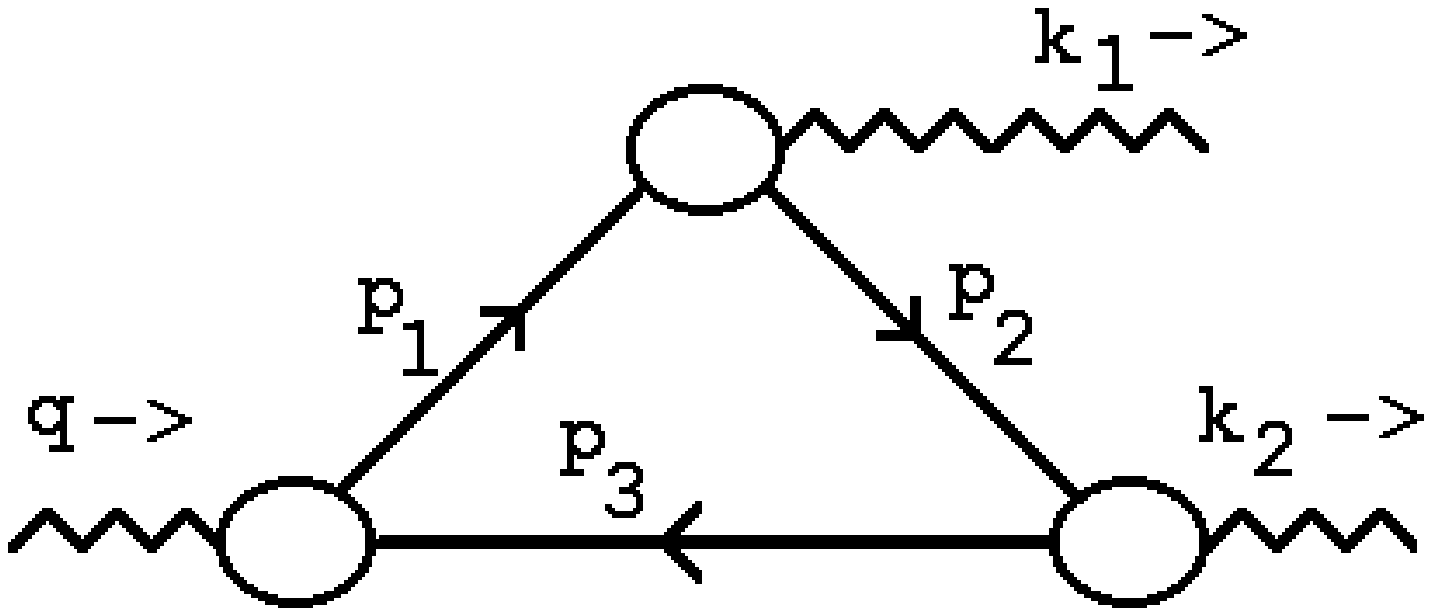}

Fig.~A.2~ The Triangle Landau Singularity - the arrows define $p_1,p_2,p_3$.
\end{center}

Multiplying (\ref{trl}) by each of the 
$p_i$ gives a set of three equations for which the determinental condition that there be a non-zero solution for the $\alpha_i$ gives the equation of the potential singularity surface as 
$$
x^2 + y^2 + z^2 + 2xyz = 1
\auto\label{trs}
$$
where
$$
 x=(\frac{q^2}{2m^2} -1)~, ~~
y=(\frac{k_1^2}{2m^2} -1)~,~~
z=(\frac{k_2^2}{2m^2} -1)
\auto\label{trs0}
$$
Setting $k_1^2 =0$ gives $y=-1$ and (\ref{trs}) becomes
$$
x^2 + z^2  - 2xz = (x-z)^2 = 0 ~, ~~~=> ~ x=z ~ => ~ q^2 = k_2^2
\auto\label{trs1}
$$
showing that if there is an ``anomaly pole'' at $q^2= k_2^2$, it should be
directly associated with the triangle Landau singularity . Notice that the mass dependence cancels in (\ref{trs1}) so that, 
{\it when $k_1^2=0$, the triangle singularity will occur at the 
anomaly pole position independently of the fermion mass.}  

On the physical sheet, the $i\epsilon$
prescription implies that there is an additional ``positive $\alpha$'' condition that must also
be satisfied for the contour to be trapped. This condition (that all the
$\alpha$'s in (\ref{trl}) must be positive) is equivalent to requiring that a physical rescattering process be possible in co-ordinate space - with all internal propagators on-shell. 
The $\alpha_i$ are then proportional to the flight time of the internal
particles. For example, for a two-particle threshold, this condition requires that particles can not have any relative transverse momentum that would lead to them traveling apart, and so be unable to rescatter. Therefore, the physical Landau singularity occurs, not surprisingly, at the threshold when the two particles are produced at rest in the center of mass frame.

The physical sheet triangle singularity occurs when, in the notation
of Fig.~5.2, an initial current carrying momentum $q$ produces two physical particles, one of which produces a current carrying momentum $k_1$ and continues as a physical particle before combining with the other particle to produce a current carrying momentum $k_2$. A priori, the initial particles could carry relative transverse momentum if the intermediate  
current interaction injects transverse momentum that reverses that of the ongoing 
particle and allows it to rescatter. This requires $k_1^2 < 0$, and so 
when $k_1^2=0$ the physical region singularity also occurs only at the threshold $q^2=4m^2$ - when the initially produced particles have no relative transverse momentum. Consequently, the physical triangle singularity can contribute at $q^2=0$ only when the internal particles are massless.

At a general value of $q^2 = k_2^2$, and with $k_1^2=0$,
the triangle singularity surface 
corresponds to mixed signs of the $\alpha$'s in (\ref{trl}), implying that at least one of the internal on-shell particles has negative energy. In this case, the singularity should occur only on unphysical sheets of the physical amplitude.
(We will enlarge on the significance of this possibility shortly.)
In triangle diagram amplitudes, the large momentum (Pauli-Villars?) pole
plays an essential role in ensuring the correct physical sheet appearance of the triangle singularity.

When $k_1^2=k_2^2$, it is obviously not possible for $k_2^2$ to be timelike while $k_1^2$ is spacelike and so the positive $\alpha$ triangle condition is again satisfied only
when the two initially produced particles have no relative momentum. It occurs, therefore, when $k_1 = k_2 =q/2$, giving $q^2 = 4 k^2$. The mass-shell condition then imposes $k^2 = m^2$
and so, again, the physical region triangle singularity occurs only at the two-particle
threshold. If we set $k_1^2 =k_2^2$ and write $y=z= - 1 + \delta$ then, for small $\delta$, the solution to (\ref{trs}) is 
$$ x = -1 ~+~ 4\delta~ +~ O(\delta^2) 
\auto\label{ts13}
$$
which implies that
$$
q^2 = ~4 k_1^2 = ~4 k_2^2
\auto\label{ts131}
$$ 
is indeed the (unphysical) triangle singularity surface close to threshold. However, since we do not have full amplitudes for the $k_1^2= k_2^2$ case, we will not discuss this kinematics any further.

\subhead{A.9 The Unphysical Triangle Singularity at Infinite Momentum}

In general, Landau singularities occur more extensively on unphysical sheets defined by analytic continuation around the physical region branch cuts.  The continuation distorts the integration contour
and, as a result, a singularity can occur on the branches of the 
singularity surface defined by the mass-shell
conditions together with a mixed set of signs for the $\alpha$'s.
This is how the triangle singularity occurs as a double pole in $A_{q^2}^6$ and
$A_{k_2^2}^6$. However, as we have seen, it also occurs as a single pole in $A^6_{\infty}$ 

When $k_1^2 =0$ and $q=q_{\perp}$ is spacelike, $k_2^2 = q^2$ implies 
$$
k_2 = q_{\perp} + k_1^+
\auto\label{trik}
$$
where $k_1^+$ is lightlike and orthogonal to $q_{\perp}$. If $p$ is the internal loop momentum then, if $p_{\perp} = - q_{\perp} / 2$, and $p^+ \to \infty$,
all three mass-shell conditions become
$$
p_i^2 ~~\centerunder{$\sim$}{${\scriptstyle p^+ \to \infty}$}~~
p^+p^-~-~q_{\perp}^2 / 4 ~=~m^2
\auto\label{trika}
$$  
and so (\ref{trsm}) and (\ref{trl}) will be satisfied asymptotically if
$$
{p}_i \sim p^+ \to \infty~, ~~p^+ \parallel k^+ ~, ~~~~\alpha_1 + \alpha_2 + \alpha_3 = 0~,
\auto\label{trs7}
$$
Therefore, if there is a non-zero contribution from the large momentum region of the integral, a triangle singularity could be generated. Because of the large momentum divergence producing the surface contribution involving the anomaly, this can indeed be the case. A pole generated in this way will have the property that (when
$k_1^2 =0$) it is independant of the fermion mass and so can survive in the infinite mass limit of the subtracted Pauli-Villars amplitude. The circulating large light-like momentum implies, however, that the propagators necessarily carry a combination of positive and negative infinite energies.

In the various limits discussed in subsection {\bf A.6}, we have shown that while the triangle singularity (anomaly pole) at $q^2 = k^2_2~$ is present in parts of the amplitude, in 
all cases, there is a cancelation such that there is no singularity on the physical sheet. As we have also discussed, it appears to be possible
for the anomaly pole terms in $A_3$ and $A_6$
to be viewed as a Pauli Villars subtraction that gives an amplitude that falls off in the infinite mass limit. 

In effect, imposing vector current conservation via a
Pauli-Villars subtraction can compensate for an incorrect 
i$\epsilon$ prescription.
If we initially use a ``wrong'' i$\epsilon$ 
prescription so that an unphysical
triangle singularity is present in the physical amplitude,
there will be a cancelation between ``unphysical'' triangle singularities generated in the original triangle amplitude and in the subtracted
unphysical Pauli-Villars amplitude. In both cases the mixed $\alpha$ condition implies that the propagators carry a combination of positive and negative energies. We will discuss how a pole can be generated in these circumstances in sub-section {\bf A.13}.

We have seen that the threshold branch-points are closely related to the infra-red triangle singularity. If the threshold singularities are eliminated {\it as a consequence of confinement} so that there is no infra-red anomaly pole, the large momentum anomaly pole could remain as a Pauli-Villars contribution to bound states. 

\subhead{A.10 The Physical Triangle Singularity}

We have already discussed 
the appearance\cite{cg} of the anomaly pole in $T^{\scriptscriptstyle AAA}$
in {\bf A.5}. How the pole is produced by the triangle singularity
can be understood\cite{cg} as follows. 
Ignoring, at first, the fermion numerators, the S-Matrix formula for the triangle discontinuity integral is
$$
I~=~ \int d^4p \prod_{r=1}^{3} \theta(p_r^0)\delta(p_r^2)
\auto\label{trI} 
$$
For any external or internal four-momentum $p$ we can define
$$
\tilde{p} ~=~ Q^{-1} \{(e_1.p)~e_1 ~+ ~(e_2.p) ~e_2\} ~+ ~(e_-.p)~e_+ ~+~
Q^{-2}~(e_+.p) ~e_-\}
\auto\label{chv}
$$
where $e_1,e_2,e_+,e_-$ are the basis vectors (\ref{bvs}). The reverse transformation is 
$$
p ~=~ Q \{(e_1.\tilde{p})~e_1 ~+ ~(e_2.\tilde{p}) ~e_2\} ~+  
~(e_-.\tilde{p})~e_+ ~+~
Q^2~(e_+.\tilde{p}) ~e_-\}
\auto\label{chvr}
$$
After the transformation (\ref{chv}), the external momenta (\ref{+lc}) become
$$
\tilde{k}_1 = e_1 + e_+,~~~ \tilde{k}_2= (-e_1 +\sqrt{3}e_2)/2, ~~~\tilde{k}_3= -(e_1 +\sqrt{3}e_2)/2
-e_+
\auto\label{+lci}
$$
and so are independent of $Q$.

Noting that
$$
d^4p~=~~Q^4~ d^4 \tilde{p}
\auto\label{chv1}
$$
and that
$$
\delta (p^2) ~=~ Q^{-2} \delta (\tilde{p}^2) ~+~ \tilde{p}^{-2} \delta (Q^2)
\auto\label{chv2}
$$
we observe that we can 
produce  $\delta (Q^2)$ from (\ref{trI}) if two internal lines, $p_1$ and $p_2$, contribute via the first term in (\ref{chv2}) while the 
$p_3$ line contributes via the second term. 
Substituting (\ref{chv1}) and (\ref{chv2}) and writing 
$\tilde{p}_3 = (\tilde{p}_3^0,~ \underline{\tilde{p}})$, (\ref{trI}) gives
as the coefficient of $\delta(Q^2)$
$$
\eqalign{&\int d^3\underline{\tilde{p}} \int \frac{d\tilde{p}_3^0 \theta(\tilde{p}^0_3)}{\tilde{p}_3^2}  ~ \delta(\tilde{p}_1^2)~
\delta(\tilde{p}_2^2)
~=~\int d^3\underline{\tilde{p}}
\int_0 \frac{d\tilde{p}_3^0 }{(\tilde{p}^0_3)^2
 - (\underline{\tilde{p}})^2} 
\delta(\tilde{p}_1^2 (\tilde{p}^0_3))~
\delta(\tilde{p}_2^2(\tilde{p}^0_3))
\cr
& =~
\int d^3\underline{\tilde{p}} ~ \frac{1}{|\underline{\tilde{p}}|}~\biggl[ln~ \biggl(~\frac{\tilde{p}^0_3 + |\underline{\tilde{p}}|}{\tilde{p}_3^0 -|\underline{\tilde{p}}|}~\biggr)\biggr]_{
\tilde{p}_3^0 = 0}  
\delta(\tilde{p}_1^2)_{\tilde{p}^0_3=0}~
\delta(\tilde{p}_2^2)_{\tilde{p}^0_3=0} ~ + ... \cr
& =~ i\pi ~\int d^3\underline{\tilde{p}}~ \frac{1}{|\underline{\tilde{p}}|}
~\delta(\tilde{p}_1^2)_{\tilde{p}^0_3=0}~
\delta(\tilde{p}_2^2)_{\tilde{p}^0_3=0}}
\auto \label{p30}
$$
For $\tilde{p}_1$ and $\tilde{p}_2$ to be lightlike with $\tilde{p}^0_3 = 0$, it 
must be that the external light-like momentum $e_+$ flows along these lines so that  $\tilde{p}_1^0 = \tilde{p}_2^0 = 1$.
$\tilde{p}_3$ is spacelike, therefore, and varies over the finite region allowed by
$(\tilde{p}_1^2)_{\tilde{p}^0_3=0}=(\tilde{p}_2^2)_{\tilde{p}^0_3=0}=0$.

Making the reverse transformation (\ref{chvr}) we find that, 
in terms of the original momenta, the pole in the $q^2$ channel is produced by (in the notation of Fig.~A.1(a)) the 
internal momentum region $p \sim 0$, with $k_1 \sim e_+$ 
and $k_2 \sim Q  \to 0$. That there is zero internal momentum on one line 
will be very significant when we discuss how the fermion numerators can produce the momentum structure of the anomaly. We first discuss the definition of chiral amplitudes in the massless limit.

\subhead{A.11 Chiral Amplitudes for $m\to 0$}

At first sight, a very simple chiral amplitude structure emerges 
when $m=0$. We introduce right and left-handed currents R and L by writing
$$
A_{\mu}~=~R_{\mu}~+~L_{\mu}, ~~~~ V_{\mu}~=~R_{\mu}~-~L_{\mu}, 
\auto\label{RL}
$$
with 
$$
R_{\mu}~=~ \frac{(1 + \gamma_5)}{2} ~\gamma_{\mu} ~...~,~~
L_{\mu}~=~\frac{(1 - \gamma_5)}{2} ~\gamma_{\mu}~ ...
\auto\label{RL1}
$$
Setting $m=0$ in the appropriate variation of (\ref{tamp}) and (naively) anti-commuting the $\gamma_5$
factors maximally, we find that chirality is conserved along the internal fermion line, implying that only the 
$T^{\scriptscriptstyle RRR}$ and $T^{\scriptscriptstyle LLL}$ amplitudes are non-zero. Moreover, 
$$
T^{\scriptscriptstyle AAA} ~=~T^{\scriptscriptstyle AVV}
~=~T^{\scriptscriptstyle VAV}~=~T^{\scriptscriptstyle VVA}
~=~T^{\scriptscriptstyle RRR}~+~T^{\scriptscriptstyle LLL}
\auto\label{RL2}
$$
However, there is now a major conflict between the axial-vector and vector
current Ward identity divergences. Obviously, if (\ref{RL2}) holds,
we can not simultaneously have vector current conservation and an anomalous axial current divergence.

Fortunately, perhaps, this conundrum is connected to the well-known difficulty 
in regularizing amplitudes containing
$\gamma_5$ in a manner that maintains the full $\gamma$-matrix algebra. Dimensional regularization (which, at first sight, would appear to give what is wanted) can not be used because there is no definition of an anticommuting $\gamma_5$ away 
from four dimensions that also satisfies  
$$
Tr\{\gamma^{\alpha}
\gamma^{\beta}\gamma^{\gamma}\gamma^{\delta} \gamma_5\}
~\neq~ 0
\auto\label{g5}
$$
and so can smoothly give the four-dimensional result
$$
Tr\{\gamma^{\alpha}
\gamma^{\beta}\gamma^{\gamma}\gamma^{\delta} \gamma_5\}
~=~ 4 i ~\epsilon^{\alpha \beta\gamma\delta}
\auto\label{g51}
$$
(\ref{g51}) is crucial in obtaining the anomaly amplitudes that we are discussing.
In general, there is no way to handle the divergence of triangle amplitudes 
that respects the full algebraic properties of $\gamma_5$. In particular, $\gamma_5$ can not be freely anticommuted around the triangle diagram, as is done to obtain
(\ref{RL2}).

We can, however, obtain massless anomaly amplitudes that satisfy vector current conservation if we use (\ref{RL}) directly and take the $m \to 0$ limit of initially massive amplitudes. Because the large momentum anomaly structure is independent of the fermion mass and, moreover, contains the contribution of infinitely massive fermions, the chirality violating amplitudes,
$T^{\scriptscriptstyle RRL},~T^{\scriptscriptstyle RLR},$~...~, actually do not vanish but instead contain the same anomalies and anomaly pole contributions
as the amplitudes that do not violate chirality. We construct
a full set of anomaly amplitudes as follows.

We start with well known property of the anomaly that it appears only in 
AAA and AVV amplitudes. Therefore, considering only anomaly contributions
to amplitudes, we can set 
$$
T^{\scriptscriptstyle AAV} ~=~T^{\scriptscriptstyle AVA}~=~...~=
T^{\scriptscriptstyle VVV} ~=~0
\auto\label{an0}
$$
Also from the foregoing, we know that both vector current conservation 
and the anomaly divergence are encapsulated in the 
anomaly pole amplitude (\ref{exa1}) which we have seen emerge explicitly in
$T^{\scriptscriptstyle AAA}$. A sum of such terms gives the 
full $T^{\scriptscriptstyle AAA}$ amplitude when $k_1^2 = k_2^2= k_3^2$ and we have
seen that, in this kinematics, a single term of this form should similarly provide the  $T^{\scriptscriptstyle AVV}$ amplitude. Therefore, in the symmetric kinematical situation $k_1^2 = k_2^2= k_3^2$ we define 
$$
T^i~=~T^i_{\alpha \beta \gamma}(k_i,k_j,k_k) ~=~ k^i_{\alpha}~\frac{
{\hbox{\large $\epsilon$}}_{\delta \sigma\beta\gamma}~ k_j^\delta 
k_k^{\sigma} }{8\pi^2 (k^i)^2}
\auto\label{Tijk}
$$
with $T^j$ and $T^k$ defined by simultaneous rotation of $(i,j,k)$ and
$(\alpha,\beta,\gamma)$. 

The set of amplitudes
$$
\eqalign{&T^{\scriptscriptstyle R_iR_jR_k}~=~T^{\scriptscriptstyle L_iL_jL_k}
~=~T^i~+~T^j~+~T^k \cr
&T^{\scriptscriptstyle R_iL_jL_k}~=~T^{\scriptscriptstyle L_iR_jR_k}~
= ~T^i \cr
&T^{\scriptscriptstyle L_iR_jL_k}~=~T^{\scriptscriptstyle R_iL_jR_k}~
= ~T^j \cr
&T^{\scriptscriptstyle L_iL_jR_k}~=~T^{\scriptscriptstyle R_iR_jL_k}~
= ~T^k }
\auto\label{TiA}
$$
then gives
$$
\eqalign{&T^{\scriptscriptstyle AAA}~=~
T^{\scriptscriptstyle RRR}+T^{\scriptscriptstyle RRL}+T^{\scriptscriptstyle RLR}+T^{\scriptscriptstyle RLL}+T^{\scriptscriptstyle LRR}+T^{\scriptscriptstyle LRL}+T^{\scriptscriptstyle LLR}+T^{\scriptscriptstyle LLL}\cr
&~~~~~~~=~4 (T^i+T^j+T^k)}
\auto\label{TijkA}
$$
together with
$$
\eqalign{&T^{\scriptscriptstyle AVV}~=~
T^{\scriptscriptstyle RRR}-T^{\scriptscriptstyle RRL}-T^{\scriptscriptstyle RLR}+T^{\scriptscriptstyle RLL}+T^{\scriptscriptstyle LRR}-T^{\scriptscriptstyle LRL}-T^{\scriptscriptstyle LLR}+T^{\scriptscriptstyle LLL}\cr
&~~~~~~~=~4 T^i }
\auto\label{TijkV}
$$
while also satisfying (\ref{an0}), e.g.
$$
\eqalign{&T^{\scriptscriptstyle AAV}~=~
T^{\scriptscriptstyle RRR}-T^{\scriptscriptstyle RRL}+T^{\scriptscriptstyle RLR}-T^{\scriptscriptstyle RLL}+T^{\scriptscriptstyle LRR} -T^{\scriptscriptstyle LRL}+T^{\scriptscriptstyle LLR}-T^{\scriptscriptstyle LLL}\cr
&~~~~~~~=~0}
\auto\label{TijkAV}
$$

In this way, we obtain a set of massless anomaly pole amplitudes that gives both the correct axial anomaly divergence and vector current conservation in all channels and is also consistent with all of the foregoing. Moreover,
when these amplitudes are extrapolated to more general kinematic
situations they must have, of course, normal physical sheet analytic
properties. We assume that this requires the appearance of the same threshold structure, infra-red anomaly poles, and large momentum anomaly pole cancelations, that we have seen in $T^{\scriptscriptstyle AVV}$.

Clearly, if we insist on conservation of the vector currents,
chirality can not be conserved.
That the chirality non-conservation appears in the triangle discontinuity, in the manner that we describe next,
is of major significance in our reggeon diagram analysis.

\subhead{A.12 Chirality Violating Anomaly Pole Amplitudes}

While it is, perhaps, not so surprising that the ultra-violet
regularization of the anomaly that conserves vector current Ward identities, necessarily violates chirality in the massless limit, it is surely
a more striking phenomenon that anomaly poles appear in both the chirality 
violating and the non chirality violating amplitudes via the triangle singularity
configuration discussed in subsection {\bf A.9}.

We begin with a straightforward discussion of the anomaly pole in 
$T^{\scriptscriptstyle AVV}$. 
The limiting momentum configuration in the triangle discontinuity is shown in Fig.~A.3, along with all the $\gamma$-matrices involved. The numerator trace that appears in the pole residue is evaluated by proceeding cyclically around the loop.
Because $Q\to 0$, a $K_{\alpha}$ factor appears in the numerator 
residue  as shown, while the remaining $\epsilon$-tensor factor, needed to 
produce a residue of the form of (\ref{Tijk}), is provided via (\ref{g51}). 
\begin{center}  
\epsfxsize=5in
\epsffile{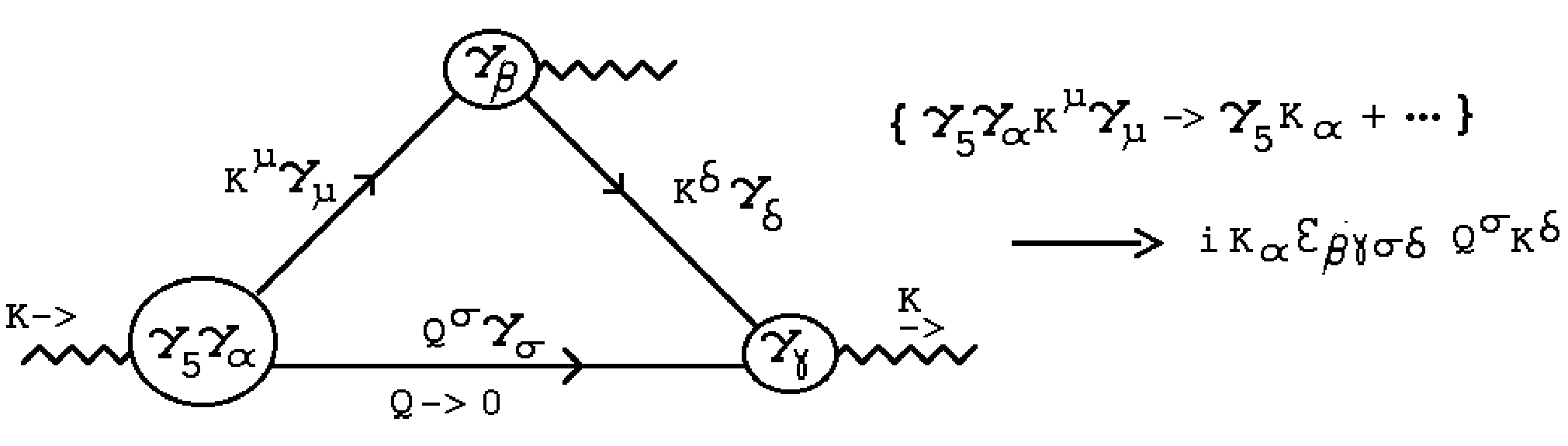}

Fig.~A.3 The Anomaly Pole Numerator in $T^{\scriptscriptstyle AVV}$ - the arrows
denote positive energy
\end{center}

A residue of the form of (\ref{Tijk}) emerges in the same way for $T^{\scriptscriptstyle RRR}$.
The result of replacing the vertices in Fig.~A3 by right-handed vertices,
and anti-commuting the $\gamma_5$ factors, is illustrated in Fig.~A4. (Since a convergent discontinuity is involved, and not a divergent feynman amplitude, there is no problem in anticommuting $\gamma_5$'s.) 
\begin{center}  
\epsfxsize=5.5in
\epsffile{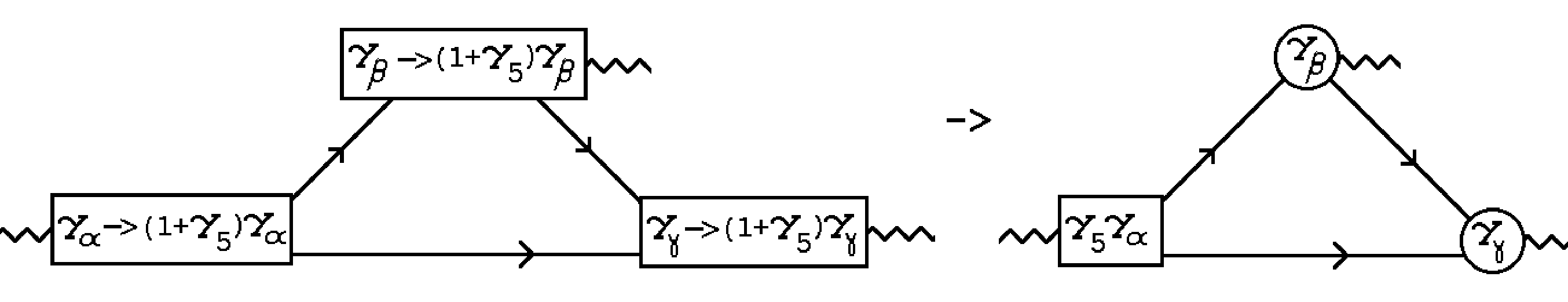}

Fig.~A.4 Vertices for the $T^{\scriptscriptstyle RRR}$ numerator - with positive energy arrows
\end{center}

That a residue of the form of (\ref{Tijk}) also emerges for the chirality violating amplitude $T^{\scriptscriptstyle LRR}$ depends much more significantly on the presence of 
the zero momentum line. As is evident from (\ref{p30}), both the antiparticle and particle poles associated with this line contribute to the emergence of $\delta(Q^2)$. As a result, it is possible for couplings to both
to be present simultaneously. If this is the case then,
in the residue at zero momentum, the vertices at the ends of
the line can both involve particle annihilation (or particle creation), as would be the result of vacuum pair production by a mass term. In effect, the zero momentum line makes a transition from a ``negative energy'' zero momentum state to a ``positive energy'' zero momentum state. The corresponding diagram is drawn as in Fig.~A.5, with the vertices shown.
\begin{center}  
\epsfxsize=3.3in
\epsffile{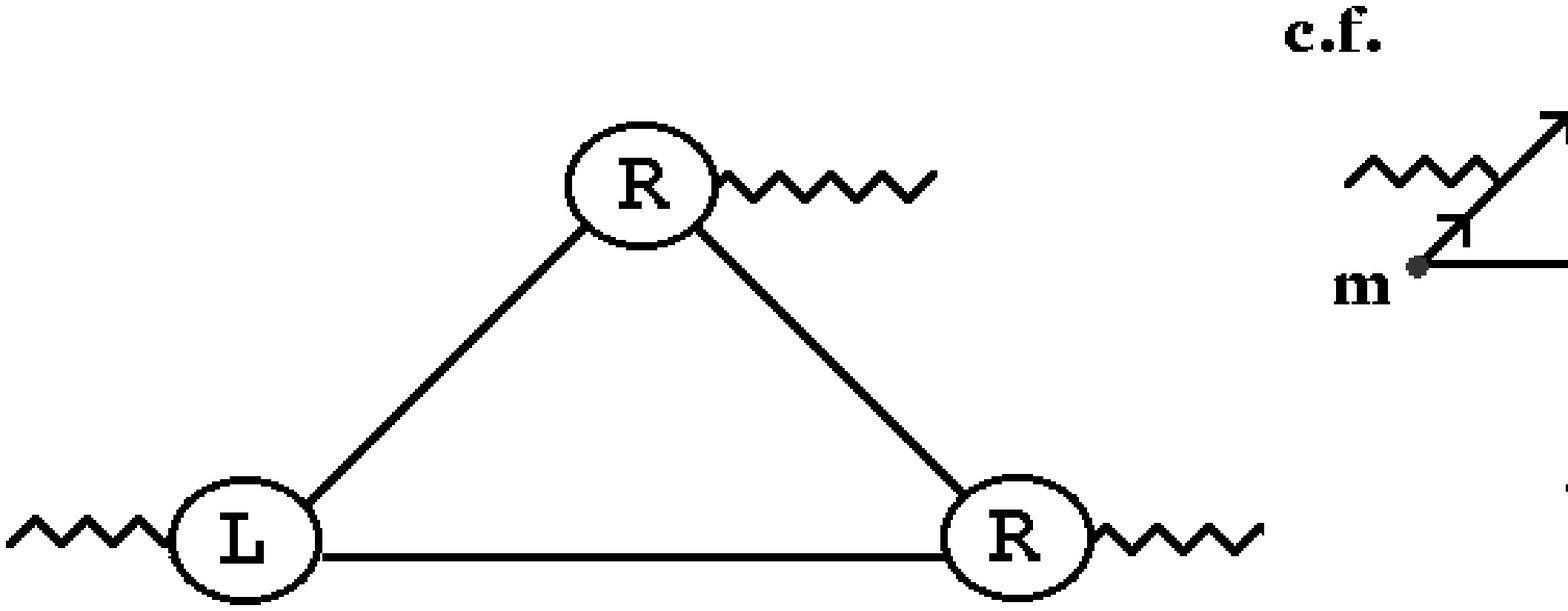}
\epsfxsize=2.4in
\epsffile{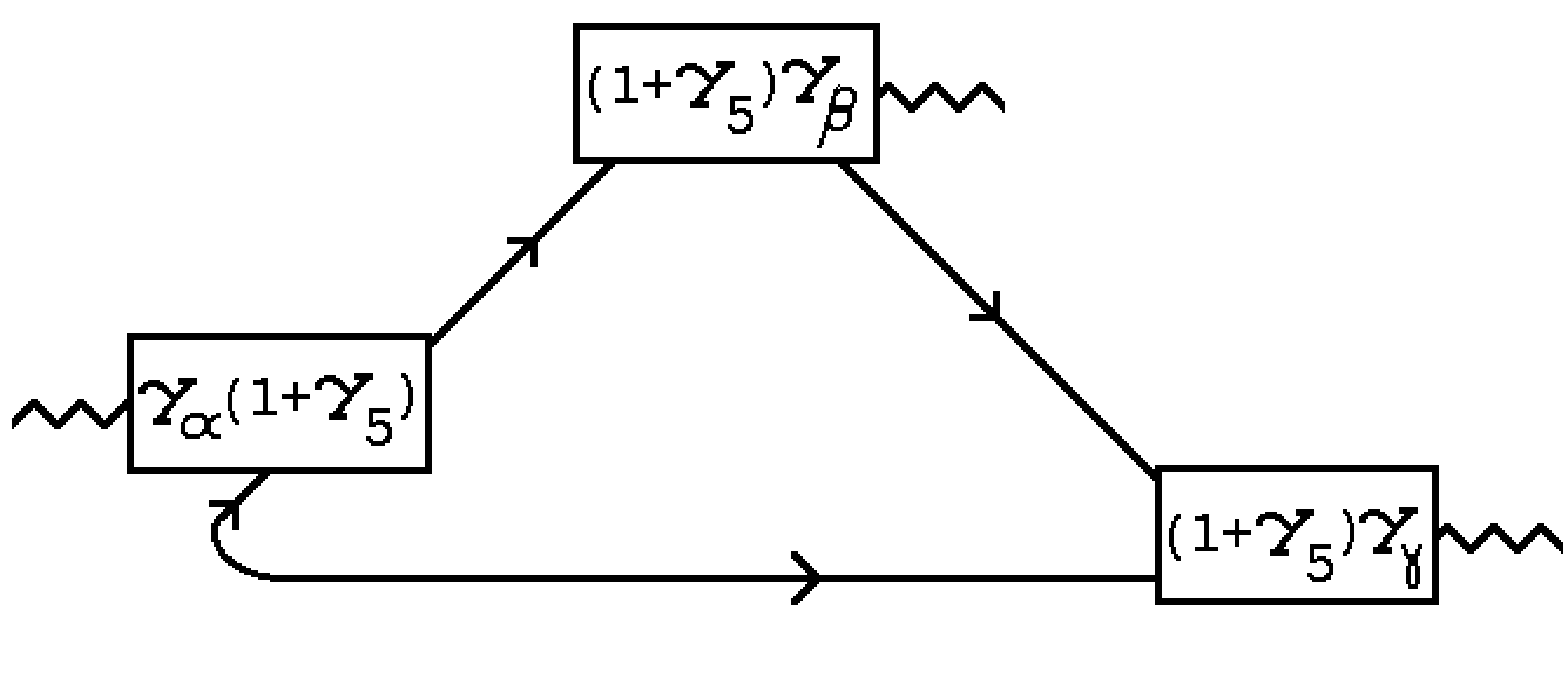}

Fig.~A.5 Vertices for the $T^{\scriptscriptstyle LRR}$ numerator - with positive energy arrows
\end{center}
Anticommuting the $\gamma_5$'s will again give a residue of the form of (\ref{Tijk}),
as anticipated in the previous subsection. 

The chirality violation along the zero momentum line is analagous to the zero momentum contribution of a propagator to a chiral condensate and so it
is clearly possible that the resulting anomaly pole could be a Goldstone boson associated with the breaking of a chiral symmetry. We will discuss the kinematical
issues associated with this possibility shortly.
A general implication of the chirality violation is, obviously, that amplitudes which would be naively expected to vanish as $m^2 \to 0$, actually need not vanish. The implications of this phenomenon will depend on the quantum numbers carried by the amplitudes involved.

Consider also the possible appearance of an anomaly pole as a large momentum effect. 
As we discussed in sub-section  {\bf A.10}, an unphysical triangle singularity is potentially generated at infinite (light-cone) momentum. A pole will be generated in a similar manner to our finite momentum discussion if it is the case that, after the anomaly is subtracted, the integral at infinity can be treated analytically by 
(effectively) compactifying momentum space so that positive and negative energies are identified at infinity. (We will not attempt to discuss this in detail, but will simply assume it can be done.)
With the external momentum directed as before then, as illustrated in Fig.~A.6,
\begin{center}  
\epsfxsize=3.5in
\epsffile{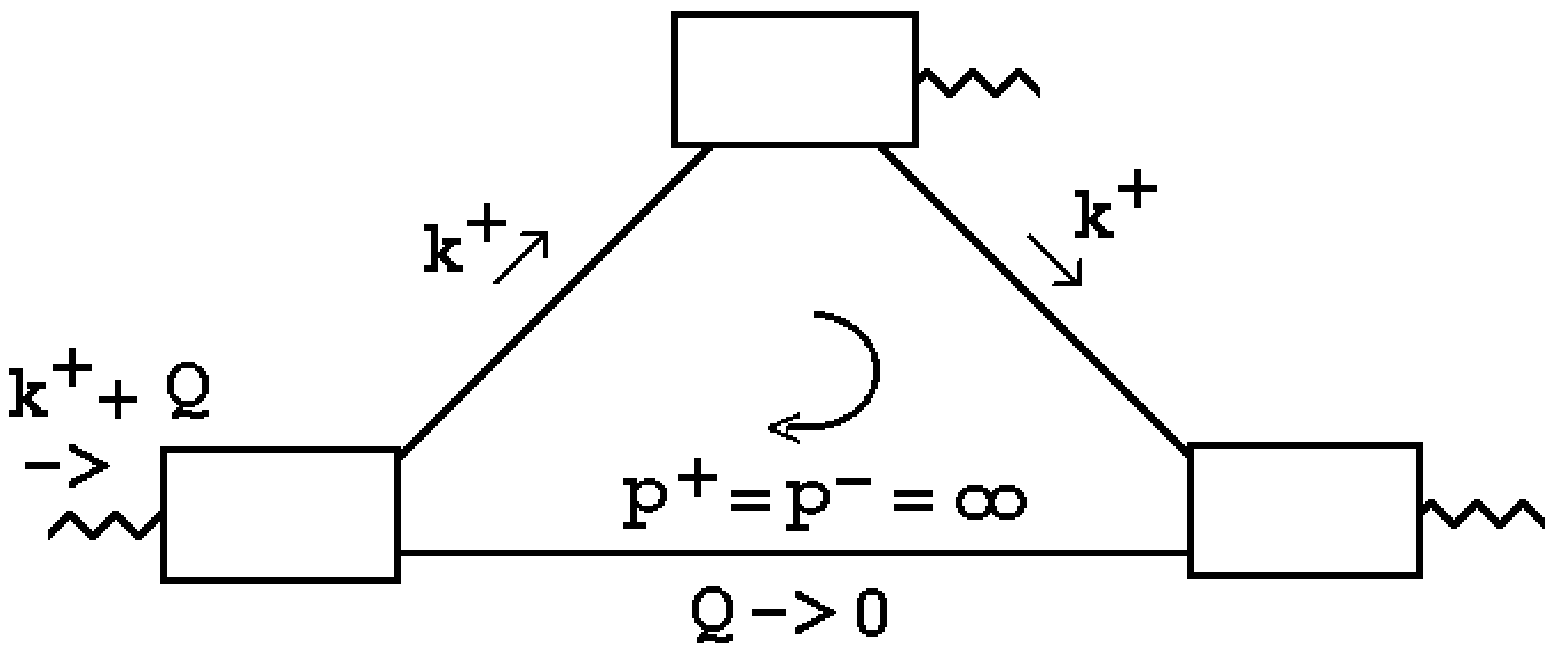}

Fig.~A.6 Momenta generating the infinite momentum anomaly pole
\end{center}
when $Q\to 0$ one line can carry infinite {\it positive~=~negative} internal momentum and so can generate an anomaly pole in parallel with the zero momentum case. Assuming that the pole residue
results from the finite momentum propagator numerators, as in the Ward identity contribution (\ref{tamp0}), then the relevant contribution, together with the corresponding vertices, will again be as in Fig.~A.3. 
The appearance of the infinite momentum pole in chiral amplitudes will
again be decribed by Figs. A.4 and A.5. Clearly, the infinite momentum line can 
involve a chirality transition that parallels the zero momentum transition. 

In parallel with our discussion at small momentum, the infinite momentum 
anomaly pole can also be viewed as an infinite light-cone momentum shift of the Dirac sea, that cancels the anomaly in the vector current. Considering the analogue of Fig.~A.5 that takes place at infinite (internal) momentum, the initial vertex involves the production of an off-shell, negative energy, fermion together with an on-shell
state that is ``produced'' via the absorption of a
negative energy state and which is then absorbed conventionally, after the helicity transition, as an on-shell positive energy state.  
This is equivalent to a flow, during the interaction, of the infinite momentum Dirac sea.

\subhead{A.12 The Anomaly Pole as a Goldstone Boson at Infinite Momentum} 

It is not immediately apparent that there are kinematical circumstances in which
the anomaly pole can appear as a Goldstone Boson particle pole in a physical amplitude. 
The anomalous wee gluon infra-red divergence described in the text 
couples via anomaly amplitudes with $k_1^2 =0$ and it is a central element
of the dynamics we describe that anomaly poles emerge and produce bound-states.
A subtle phenomenon is involved, as we now discuss.

If we keep just the anomaly pole contributions 
of $A_3$ and $A_6$ to $T_{\mu \alpha \beta}$ we obtain
$$
T_{\mu \alpha \beta}(k_1,k_2) ~=~-~{1 \over 2 {\pi}^2 } ~
{({\hbox{\Large $\epsilon$}}_{\delta \sigma\alpha\mu}k_{1\beta}
 ~-~{\hbox{\Large $\epsilon$}}_{\delta \sigma\beta\mu} 
~ k_{2\alpha})~ k_1^{\delta}
k_2^{\sigma} \over (k_1+k_2)^2} ~~~ +~\cdots 
\auto\label{pipo}
$$
Not only does this expression not have the factorization properties expected of a
particle pole,  it also, as we have seen, does not satisfy the vector Ward identities and does not have the axial-vector anomaly. 

When $k_1^2, k_2^2 , q^2 \to 0$ we can use(\ref{epid}), as we did in obtaining (\ref{indes}), to rewrite (\ref{pipo}) as 
$$
T_{\mu \alpha \beta}(k_1,k_2) ~=~{1 \over 2 {\pi}^2 }~ \frac{
{\hbox{\Large $\epsilon$}}_{\delta\sigma\alpha\beta} [k_1 + k_2]_{\mu}k_1^{\delta}
k_2^{\sigma} }{(k_1 + k_2)^2}
~-~\frac{1}{2\pi^2}\frac{({\hbox{\Large $\epsilon$}}_{\delta \sigma\beta\mu}k_{1\alpha}
 -{\hbox{\Large $\epsilon$}}_{\delta \sigma\alpha\mu} 
k_{2\beta}) k_1^{\delta}
k_2^{\sigma}}{ (k_1+k_2)^2~} ~~~+~\cdots 
\auto\label{pipo1}
$$
where the omitted terms are less singular as 
$q^2 \to 0$. The first term has the appropriate factorised form to
provide a particle pole coupled to the axial current $A_{\mu}$, while 
the second term corresponds to $A_4$ and $A_5$ contributions in (\ref{inde}) which, as we have previously noted, do not 
contribute\cite{arw02} to reggeon anomaly vertices. 
Therefore, if a physical Goldstone boson can be isolated in reggeon amplitudes, we should be able to use the first term (which is, of course, the same as (\ref{indes}) and (\ref{indes+}) - on which we have already focussed much of our attention) to obtain it's couplings.
In this case, $[k_1 +k_2]_{\mu}$
provides the coupling to the axial current $A_{\mu}$ while the factor 
$$
{\hbox{\Large $\epsilon$}}_{\delta\sigma\alpha\beta}k_1^{\delta} k_2^{\sigma}
\auto\label{gbc}
$$
provides the coupling to currents $V_{\alpha}$ and $V_{\beta}$ when 
$ k_1^2,~k_2^2~,~ q^2~ \to~ 0$. 

There is an obvious problem, however, in using (\ref{gbc}) only in the limit in which
all three of $k_1^2~,k_2^2,$ and $q^2$ are taken to zero. In this limit
$k_1 \parallel k_2 \parallel k_+$ where $k_+$ is light-like and so, 
because of the {\Large $\epsilon$}-tensor,
(\ref{gbc}) gives a vanishing pole residue. This is hardly surprising since,
as a matter of principle, we would not expect to be able to determine the finite momentum coupling of a Goldstone boson meson, say, to vector gluons in QCD. 

Also, as a matter of principle, we would expect that the anomaly pole should only emerge as a physical Goldstone boson if it is generated in the infra-red integration region.
As described above, to see the anomaly pole emerge as an infra-red effect
when taking $k_1^2 \to 0$ we must approach the zero momentum limit by also taking 
$k_2^2 \to q^2$. If $q$ is timelike, then $k_2^2 = q^2 \implies k_1 = k_1^+ = 0$
and so the residue, which will again be given by (\ref{gbc}), will vanish directly as $k_1^+ \to 0$, {\it unless we simultaneously take an ``infinite momentum'' limit.} To avoid this, the limit must be taken with $k_1^+ \to 0$. $ k_2^- \to \infty$, such that $k_1^+ k_2^-$ remains finite, and with a spacelike component
$q_{\perp}$ included such that 
$q^2 =(k_1 + k_2)^2 - q_{\perp}^2 \to 0~$. This can be done within
the multi-regge limit discussed in the main text. In this case, 
the physical Goldstone ``pions'' are built from fermion reggeons that, in an appropriate frame, carry the infinite momentum $k_2^-$.
The light-cone momentum ($k_1^+$) that is taken to zero is carried by the wee partons that are playing a vacuum-like role.
In this situation, physical couplings of Goldstone bosons to gauge field configurations are possible, ``as a matter of principle''. 
Note that, to obtain the desired residue (\ref{gbc}), the polarizations of the wee partons (given by $V_{\alpha}$) and the physical fermions (given by $V_{\beta}$) must be orthogonal. 

\subhead{A.13 A Large $k_{\perp}$ Cut-Off, Ward Identities, and Anomaly Poles}

Finally, we discuss how a cut-off affects the foregoing discussion. An essential element of our construction of reggeon diagrams and interactions,
for both QCD$_S$ and QUD, is that we initially impose a transverse momentum cut-off that is also imposed in the anomaly interactions that we utilise. In previous papers, we have discussed the importance of this cut-off at length. In principle, we would like to view it as ensuring the validity of a $t$-channel unitarity
derivation\cite{arwt} of both reggeon exchanges (via wrong-signature nonsense states)
and reggeon interactions (due to right-signature nonsense states). However, to derive all the reggeon interactions that we discuss, particularly the anomaly
diagrams, in this way would be a major project. In practise, 
we will not discuss how the cut-off is imposed and will, in
effect, assume that the outcome is independent of any possible variation.

From the discussion in {\bf A.3}, it is clear that a cut-off in the triangle diagram will remove the surface anomaly term and will also interfere with the derivation
of the vector Ward identities. In effect, the cut-off provides an alternative regularization which necessarily violates the vector Ward identities and also modifies the axial current divergence. Crucially, for our purposes, 
because the anomaly pole is generated at finite transverse momentum, the infra-red structure will remain unchanged and so if we take the zero mass limit, together with the appropriate zero momentum limits, anomaly poles will appear in the
chiral amplitudes via (\ref{TiA}), (\ref{TijkA}), and (\ref{TijkV}),
as infra-red contributions.  

When the infinite cut-off limit is taken,
Pauli-Vilars regularization must, of course, be reintroduced to obtain
finite amplitudes.

\end{document}